\documentclass[aps, prl, twocolumn, 10pt, nofootinbib]{revtex4-1}
\usepackage[sort&compress]{natbib}
\usepackage{amsmath,amssymb,amsfonts}
\usepackage{graphicx}
\usepackage{xcolor}
\usepackage[colorlinks=true]{hyperref}
\usepackage[UKenglish]{babel}
\usepackage[UKenglish]{isodate}
\usepackage{multirow}
\usepackage{braket}
\usepackage{enumitem}
\usepackage{dblfloatfix}
\usepackage{pdfpages}
\usepackage{etoolbox}
\makeatletter
\patchcmd{\@outputpage@head}{\@ifx{\LS@rot\@undefined}{}{\LS@rot}}{}{}{}
\makeatother
\definecolor{citeblue}{HTML}{0E5484}
\hypersetup{colorlinks=true,citecolor=citeblue}
\begin{document}
\title{Room Temperature Atomic Frequency Comb Memory for Light}
\author{D. Main$^1$}
\author{T. M. Hird$^{1,2}$}
\author{S. Gao$^{1}$}
\author{I. A. Walmsley$^{1,3}$}
\author{P. M. Ledingham$^{1,4}$}
\email{P.Ledingham@soton.ac.uk}
\affiliation{$^1$Clarendon Laboratory, University of Oxford, Parks Road, Oxford, OX1 3PU, UK\\
$^2$Department of Physics and Astronomy, University College London, London WC1E 6BT, UK\\
$^3$QOLS, Department of Physics, Imperial College London, London SW7 2BW, UK\\
$^4$Department of Physics and Astronomy, University of Southampton, Southampton SO17 1BJ, UK}
\date{\today}
\begin{abstract}
We demonstrate coherent storage and retrieval of pulsed light using the atomic frequency comb quantum memory protocol in a room temperature alkali vapour. We utilise velocity-selective optical pumping to prepare multiple velocity classes in the $F=4$ hyperfine ground state of caesium. The frequency spacing of the classes is chosen to coincide with the $F'=4 - F'=5$ hyperfine splitting of the $6^2$P$_{3/2}$ excited state resulting in a broadband periodic absorbing structure consisting of two usually Doppler-broadened optical transitions. Weak coherent states of duration $2\,\mathrm{ns}$ are mapped into this atomic frequency comb with pre-programmed recall times of $8\,\mathrm{ns}$ and $12\,\mathrm{ns}$, with multi-temporal mode storage and recall demonstrated. Utilising two transitions in the comb leads to an additional interference effect upon rephasing that enhances the recall efficiency.
\end{abstract}

\maketitle

The atomic frequency comb (AFC) protocol \cite{Afzelius2009} is one of the leading quantum memory schemes to date and has been demonstrated in a number of configurations and materials \cite{Bussieres2013a, Heshami2016}. AFC  involes the absorption of a pulse of light of duration $\tau_p$ into an ensemble of two-level atoms, in which the atoms have been arranged such that the absorption spectrum consists of a series of $M$ absorbing peaks periodically spaced with a frequency separation of $\Delta$ and width $\gamma$. A collective coherence is established between the ground state  $|G\rangle$ and excited state $|E\rangle$ of the form $ \frac{1}{\sqrt{N}} \sum_{j=1}^Ne^{i\delta_jt}e^{-i\vec{k}_p\cdot\vec{z}_j}\ket{g_1,\dots,e_j,\dots,g_N}$ where  $g(e)_j$ labels the ground (excited) state of the $j^\mathrm{th}$ atom,  $N$ is the total number of atoms, $\vec{k}_p$ is the input photon wave vector, $\vec{z}_j$ is the $j^\mathrm{th}$ atom position and $\delta_j$ is the detuning of the $j^\mathrm{th}$ atom with respect to the input frequency field. Each term in this state acquires a different phase $\delta_j t$, such  that the collective dipole moment of the ensemble quickly decays. For $\gamma \ll \Delta$, the detuning is approximately $\delta_j \approx m_j\Delta$, where $m_j$ are integers with the total number of $m_j$ being the number of absorbing peaks $M$. At time $\tau = 2\pi/\Delta$, the phases of each component of the collective state are equal (to within $2n\pi$) and coherent re-emission of the light in the forward direction occurs \cite{Afzelius2009}. An important feature of the AFC memory is its modal capacity can be made arbitrarily large simply by adding more teeth $M$ to the comb without changing the optical depth \cite{Nunn2008, Afzelius2009}. The capacity to store and retrieve multimode light is a key functionality for a quantum memory to enable speedups in entanglement generation by several orders of magnitude \cite{Simon2007}.

The capacity of the AFC memory to operate in the quantum regime has been demonstrated in a number of experiments including: storage of qubits encoded in polarisation \cite{Gundogan2012, Zhou2012, Clausen2012}, time-bin \cite{Gundogan2015} and orbital angular momentum \cite{Hua2019}; storage of single photons from parametric down conversion sources \cite{Saglamyurek2011, DeRiedmatten2008, Clausen2012, Rielander2014} and quantum dot sources \cite{Tang2015}; storage of photonic entanglement \cite{Tiranov2016} and hyper-entangled states \cite{Tiranov15}; generation of photonic entanglement \cite{Kutluer2017, Laplane2017}. Large multimode capacity has been also  been demonstrated \cite{Usmani2010, Seri2019}. Other applications include the optical detection of ultrasound \cite{McAuslan2012}.

So far, these impressive experiments have utilised cryogenically-cooled rare-earth-ion-doped solids. Their holeburning properties allow the preparation of the crystal-field-induced inhomogeneous line into a series of absorbing peaks. The long-lived ground states allow ample time to prepare the relevant populations, as well as to perform the storage and retrieval. The range of applications of AFC may be increased if it were implemented in a non-cryogenic platform, as well as in systems that would allow large optical bandwidths and high optical depths. Here we show the first demonstration of the AFC protocol implemented on a room-temperature alkali vapour. We perform velocity-selective pumping on the Doppler-broadened caesium D1 line to create AFCs which we utilise to store weak coherent states of light with a duration of $2\,\mathrm{ns}$ for times up to $12\,\mathrm{ns}$. We demonstrate multimodality with storage of two temporal time-bin modes.

In our approach we utilise a room temperature caesium vapour. The D lines have relatively large transition dipole moments ($\sim\,10^{-29}\,\mathrm{Cm}$) allowing for strong light-matter coupling. Together with the relatively high vapour pressure of Caesium results in optical depths on the order of $\mathrm{OD}\,=\,1$ at room temperature, and it is straightforward to heat a Caesium cell to achieve $\mathrm{OD}\,\sim\,1000$ \cite{Thomas2017}. The simplicity in achieving high optical depth has driven successful quantum memory demonstrations using the electromagnetically induced transparency protocol \cite{Wolters2017,Namazi2017}, gradient echo memory \cite{Hosseini2011} and off-resonant Raman protocols \cite{Reim2011, Thomas2019}. A general characteristic of atoms in a vapour is the inherent Doppler effect modifying the perceived optical transition frequency of an atom depending on its velocity -- $\omega_0\,(1\,\pm\,v/c)$ -- where $\omega_0$ is the transition frequency at rest and $v(c)$ is the velocity of the atom (light).The atom velocities will be distributed according to a Maxwell–Boltzmann distribution, and the optical transition is broadened with a Doppler width of about $1\,\mathrm{GHz}$ at room temperature \cite{Preston1996}, far exceeding that of the natural linewidth of the atoms $\sim\,2\pi\,\times \,5.2\,\mathrm{MHz}$. If one maps a broadband pulse into an atomic coherence on the Doppler-broadened line, the collective polarization will dephase at a rate of $\Delta k v$, where $\Delta k$ represents the spread of $k$ vectors associated with the mapping, much more quickly than that of spontaneous emission or inter-atomic collisions.

\begin{figure}
\includegraphics[width=\linewidth]{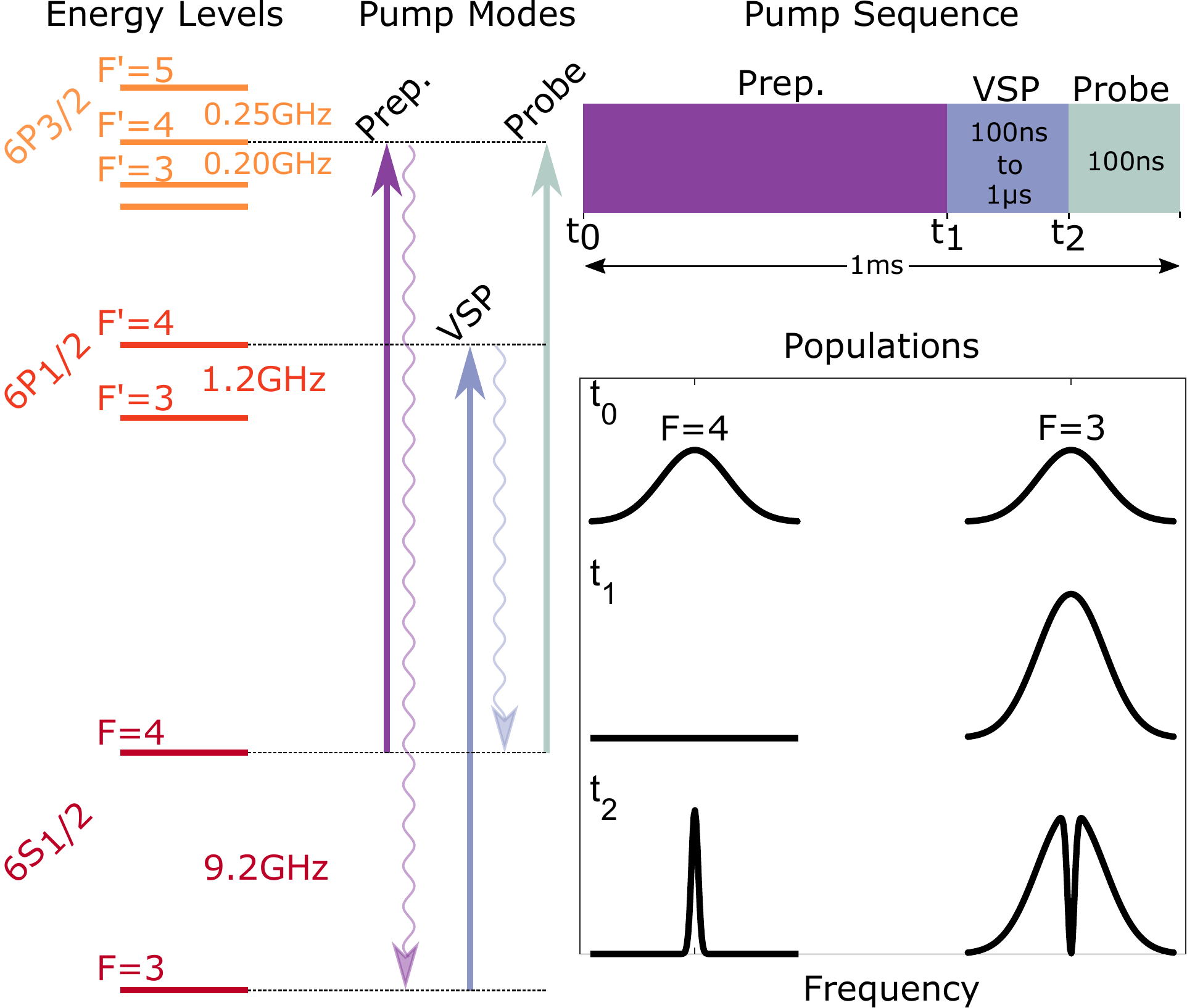}
\caption{Velocity-selective pumping principle. \textit{Left}: We show the caesium hyperfine levels for the $6\mathrm{S}_{1/2}$ ground and $6\mathrm{P}_{1/2}$, $6\mathrm{P}_{3/2}$ excited states together with the corresponding pump modes. The ``Prep.'' mode initialises the $F=4$ ground state via the  $6\mathrm{P}_{3/2}$ excited state. The ``VSP'' (velocity-selective pump) mode prepares the velocity selected population via the $6\mathrm{P}_{1/2}$ excited state. Finally, the ``Probe'' mode either probes the spectral or temporal response of the prepared feature via the $6\mathrm{P}_{3/2}$ excited state. \textit{Right}: We indicate the timescales associated with the pump sequence as well as a qualitative indication of the ground state population velocity distribution. $t_0$ represents the initial umpumped ground states. Then we apply the ``Prep.'' mode until a time $t_1$ to initialise the $F=4$ ground state and pump all population into the $F=3$. The velocity-selective pump ``VSP'' mode is then applied until a time of $t_2$ to transfer only a narrow section of the velocity distribution back from the $F=3$ state into the $F=4$ state.\label{Scheme}}
\end{figure}

The AFC memory maps naturally to this system as it makes use of inhomogenous broadening. The key step in AFC is to prepare the broadened line into a series of periodic peaks - this is achieved in an atomic vapour using velocity-selective optical pumping. This technique has been employed in alkali vapours for applications in high-resolution spectroscopy \cite{Marian2004, Aumiler2005, Ban2006, Vujicic2007},  narrowband atomic-line frequency filters \cite{Cere2009}, and steep atomic dispersion engineering \cite{Akulshin08}. Velocity-selective pumping addresses only a narrow part of the Doppler-broadened line, or in other words, a particular velocity class. Different classes can be addressed depending on the centre frequency of the pump laser, for example, the class associated with the atoms with velocity vector perpendicular to the direction of the optical mode such that $\Delta k v = 0$ corresponds to the ``zero-velocity'' class. The width of a class is determined by that of the pump, the limit being the homogeneous linewidth. If the pumping light consists of several narrowband frequency modes, then several velocity classes will be addressed and an AFC can be prepared. An alternate approach involves mapping an optical frequency comb from a mode-locked laser to the Doppler-broadened line to create atomic frequency comb structures \cite{Aumiler2005} with a tooth spacing given by the laser repetition-rate. We note that our approach has greater flexibility in the choice of tooth spacing.

Figure \ref{Scheme} shows the pumping sequence. With a paraffin-coated caesium cell of length $25\,\mathrm{mm}$ and diameter $10\,\mathrm{mm}$, we begin by completely optically pumping the $F=4$ hyperfine ground state using a laser of power around $20\,\mathrm{mW}$, duration $248\,\mu\mathrm{s}$ and waist $2.7\,\mathrm{mm}$ that is resonant with the caesium D2 transition (Sacher Lasertechnik  - the ``Prep." laser). The coating plays a key roll, allowing for thousands of cell wall collisions which shuffle the atomic velocity distribution without depolarising the spin \cite{Li2017a,Li2017b}. Combined with the high pump power giving rise to power broadening, this ensures efficient pumping of the entire Doppler-broadened line. We then use a narrowband ($\sim\,\mathrm{MHz}$) laser with waist size $1.9\,\mathrm{mm}$ that is resonant with the caesium D1 line (Toptica DFB - the ``VSP'' laser). We use this transition as the excited-state hyperfine splitting is resolved beyond the Doppler ensuring that a single velocity class from a single transition, in this case the $F=3$ to $F'=4$, is addressed. By modulating the laser drive current at particular radio frequencies (Agilent RF signal generator) we can impart frequency sidebands on the spectrum of the laser resulting in two additional narrowband frequency modes per modulation frequency, thereby enabling multiple velocity classes to be addressed with this laser. Population from these velocity classes are then pumped back into the $F=4$ hyperfine ground state with a separation $\Delta = \omega_\mathrm{mod}$. We adjust the duration of this velocity-selective pump to be faster than the atom transfer time across the spatial mode, between $100\,\mathrm{ns}$ and $1\,\mu\mathrm{s}$.

\begin{figure} 
\includegraphics[width=\linewidth]{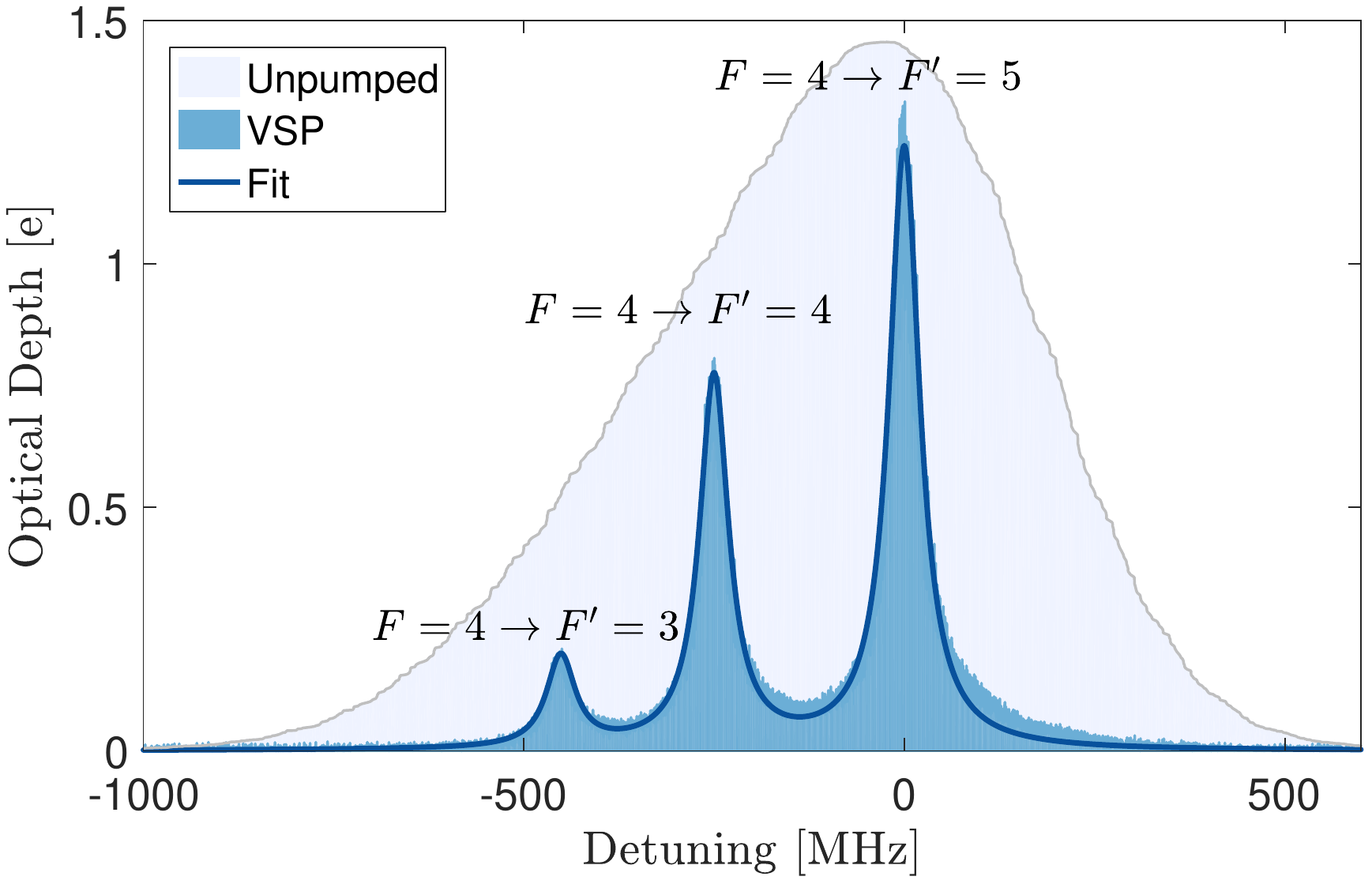}
\caption{Velocity selective optical pumping. A single velocity class is pumped back into the $F=4$ hyperfine ground state of a caesium vapour. The zero of the frequency axis represents the $F=4$ to $F'=5$ transition. The unpumped case is included for comparison. A fit is made to the features giving a width of about $45\,\mathrm{MHz}$. \label{VSP}}
\end{figure}

We probe the resulting structure using a laser on the D2 line (Toptica DL Pro - the ``Probe'' laser). This laser has a waist $998\,\mu\mathrm{m}$, spatially much smaller than the pump mode size thereby ensuring a uniform probing of the optical pumping. The probe power is kept well below the saturation intensity and has a pulse duration of around $100\,\mathrm{ns}$ to ensure that it does not induce re-pumping. We set the probe laser frequency scanning at $5\,\mathrm{Hz}$ and asynchronously to the pump sequence. This ensures that at each probe pulse has a different frequency and thus probes a different part of the spectrum. We piece together over $50$ scans to form the final spectrum. All three lasers are time gated using acousto-optic modulators (AOMs). The polarization of the probe mode is vertical and orthogonal to the counter-propagating pump modes. Experimental details are in the supplemental information. 

Figure \ref{VSP} shows an example of preparing a single velocity class in the $F=4$ ground state. The velocity-selective pump power and time are $0.86\,\mathrm{mW}$ and $1.2\,\mu\mathrm{s}$ respectively. We see three distinct peaks in the spectrum, corresponding to the three allowed transitions between the ground and excited states associated with the narrow velocity class. The spectral width of the velocity class is determined to be $45,\mathrm{MHz}$ by fitting using Gaussian functions centered at the absorption peaks. The dependence of the width on pump duration and power is described in a separate publication \cite{Main2020}. 

\begin{figure}
\includegraphics[width=\linewidth]{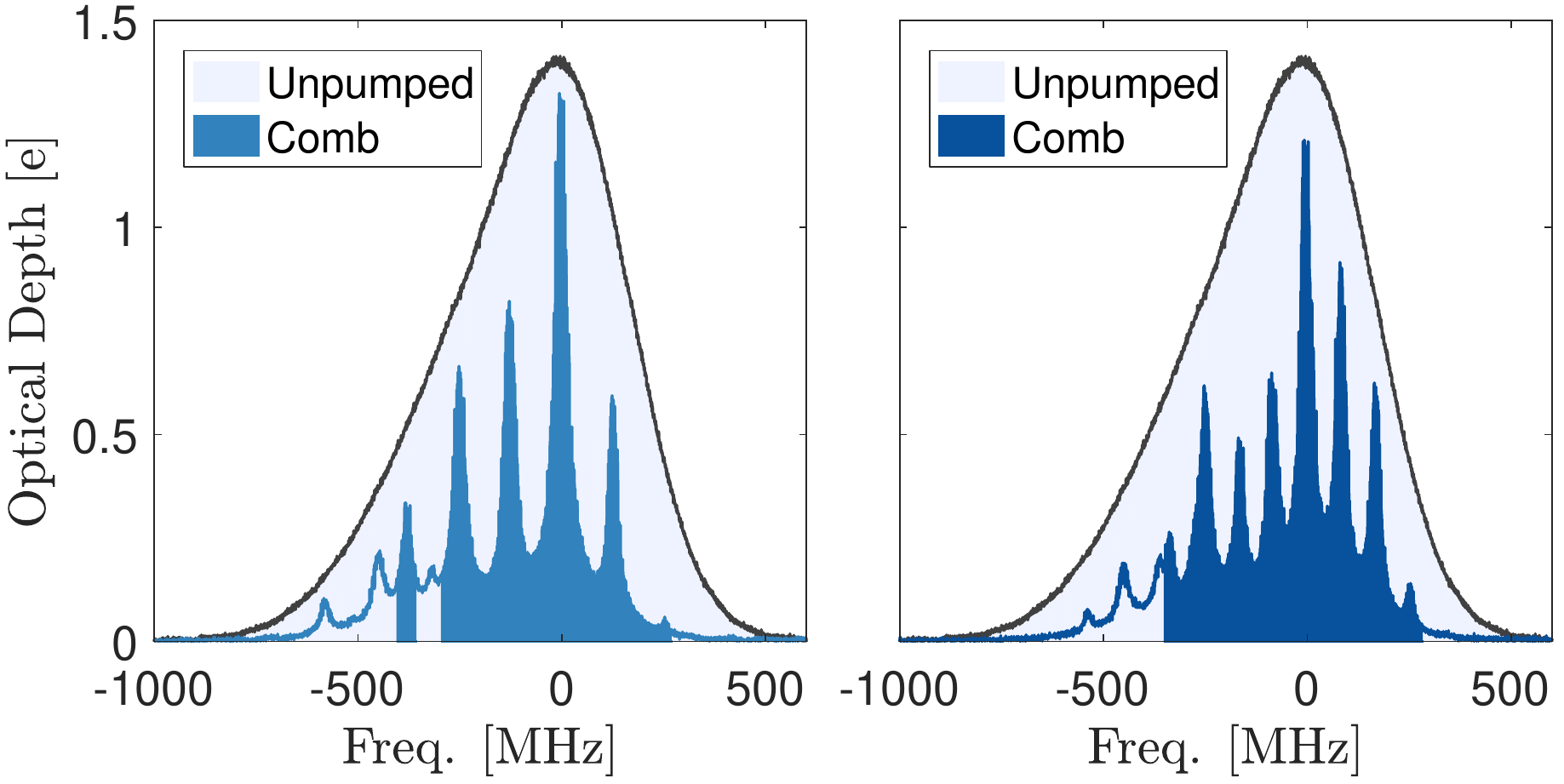}
\caption{Atomic frequency comb spectra prepared in the $F=4$ hyperfine ground state of caesium. The left panel shows the spectrum for a tooth spacing of $\Delta = 125.5\,\mathrm{MHz}$ while the right panel shows the case for $\Delta = 83.7\,\mathrm{MHz}$. The unpumped $F=4$ hyperfine ground state is also shown in both panels. The peaks that constitute the AFC are highlighted, indicating a bandwidth of around $0.6\,\mathrm{GHz}$ in both cases. We note that for the right panel, two modulation frequencies are driving the pumping diode laser current in order to prepare this comb. A detailed explanation is in the supplemental information.\label{combs}}
\end{figure}

Figure \ref{combs} shows the AFC structure. With an excited state hyperfine splitting of $251\,\mathrm{MHz}$ between the $F'=4$ and $F'=5$, we chose a comb spacing of $\Delta = 125.5\,\mathrm{MHz}$ and $\Delta = 83.7\,\mathrm{MHz}$ which is the splitting divided by $2$ and $3$ respectively. The former involves modulating the velocity-selective pump laser with a frequency $125.5\,\mathrm{MHz}$, while the latter has two modulation frequencies at $83.7\,\mathrm{MHz}$ and $167.4\,\mathrm{MHz}$. This allows us to build a larger spanning comb based on the two transitions from the $F=4$ ground state to both the $F'=4$ and $F'=5$, resulting in combs that have an acceptance bandwidth of around $0.6\,\mathrm{GHz}$. We note that this memory is therefore suitable for single photons emitted by semiconductor quantum dot sources \cite{Wolters2017}. The velocity-selective pump power and time are optimised for each case, with the peak optical depth at about $\mathrm{OD}=1$ with the comb shape roughly taking that of the initial Doppler-broadened ensemble.

\begin{figure}
\includegraphics[width=\linewidth]{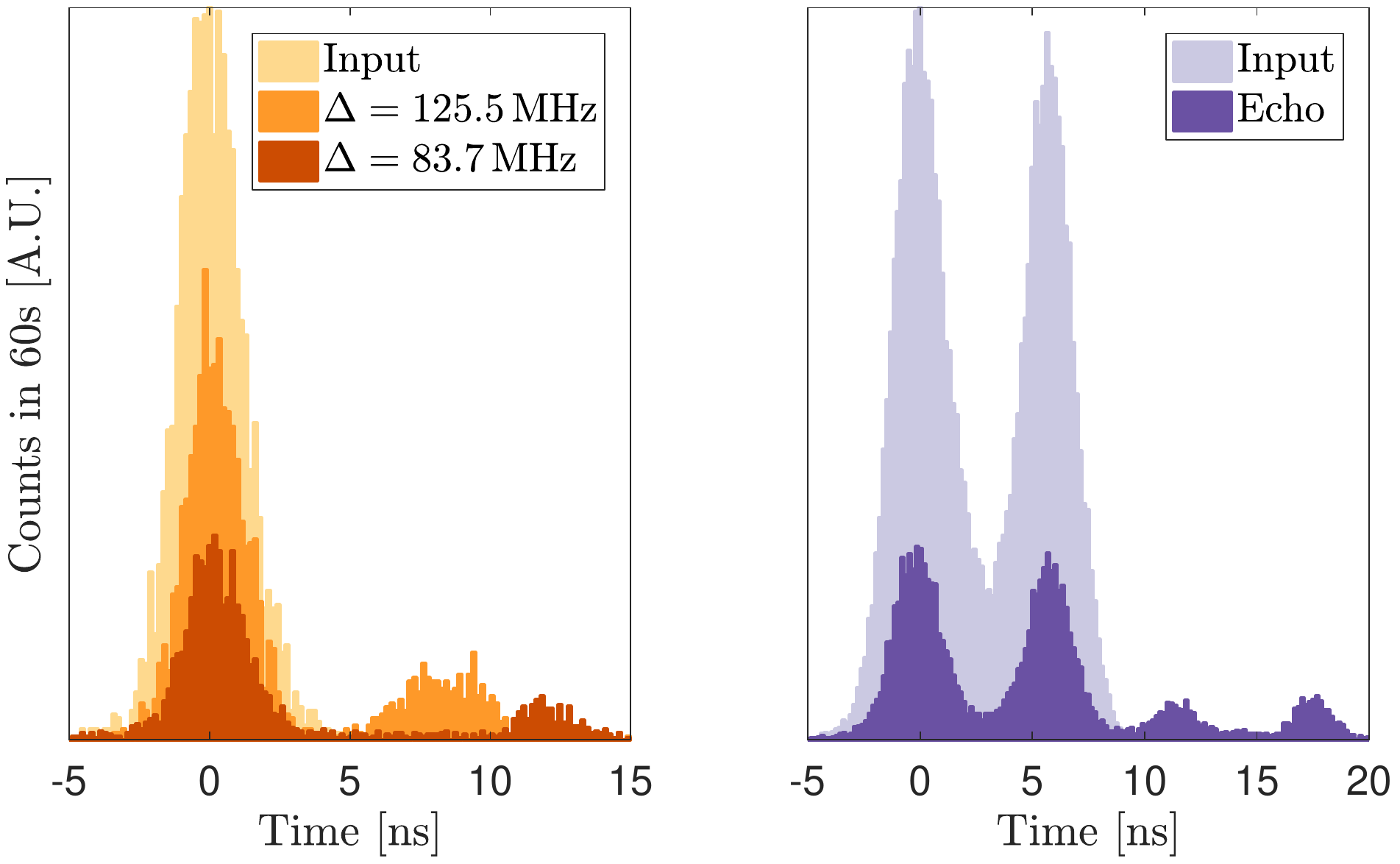}
\caption{(a) AFC storage in an atomic vapour. Two cases are shown for the two combs used in the previous figure. A clear AFC echo is seen at a storage time of $7.97\,\mathrm{ns}$ ($11.95\,\mathrm{ns}$) for the $\Delta = 125.5\,\mathrm{MHz}$ ($\Delta = 83.7\,\mathrm{MHz}$) comb. To assess the efficiency, a $5\,\mathrm{ns}$ integration window is used, centred on the relevant temporal mode. (b) Temporal multimode storage. Two temporal modes of duration $2\,\mathrm{ns}$ and separation $6\,\mathrm{ns}$ are stored in the vapour AFC for a duration of $11.95\,\mathrm{ns}$.\label{echos}}
\end{figure}

We now investigate the AFC temporal response. An input signal pulse is carved from the CW probe laser using a fibre-integrated electro-optic interferometer driven by two Tektronix arbitrary waveform generators \cite{Thomas2019}. We prepare $2\,\mathrm{ns}$ duration pulses with an approximate Gaussian shape and the intensity is reduced to a few thousands of photons per pulse at the memory with neutral density filters. We use single photon avalanche modules behind a double-pass AOM to time gate only the input pulses as well as neutral density filters to protect the detectors from preparation light leakage. Figure \ref{echos}(a) shows the time traces. For the $\Delta = 125.5\,\mathrm{MHz}$ ($\Delta = 83.7\,\mathrm{MHz}$) AFC we observe an AFC echo at the expected $7.97\,\mathrm{ns}$ ($11.95\,\mathrm{ns}$) and an efficiency of $\eta_\mathrm{AFC} = (9.3\pm 0.6)\%$ ($\eta_\mathrm{AFC} = (3.4\pm 0.3)\%$). The efficiency is measured by taking the area of the echo and comparing this to the area of the input pulse when no velocity-selective pumping has been applied. To demonstrate the multimode capability of the AFC, we prepare two input temporal modes of width $2\,\mathrm{ns}$ and separated by  $6\,\mathrm{ns}$, and clearly observe two rephased time-bin modes on the output at a storage time of $11.95\,\mathrm{ns}$ as seen in figure \ref{echos}(b). We note that the efficiency is consistent with the single temporal mode case. These results constitute what we believe to be the first implementation of an AFC optical memory in warm atomic vapour.

An expression for the AFC efficiency can be derived as in \cite{Afzelius2009} assuming Gaussian shaped peaks of equal absorption $d$ on top of a flat absorbing background $d_0$: $\eta_\mathrm{AFC} = (d/F)^2\,\mathrm{exp}[-d/F]\,\mathrm{exp}[-7/F^2]\,\mathrm{exp}[-d_0]$ where $F$ is the finesse of the comb given as $\Delta/\gamma$. As the transmitted and echo signals both propagate the same loss $\mathrm{exp}[-d_0]$, forming the ratio of the transmitted pulse to the AFC echo divides out this term and an estimate of the $d$ can be made for a given $F$. A subsequent estimate of the loss term can then be made. With a tooth width of around $45\,\mathrm{MHz}$ leading to a finesse of around $F=1.9$ for the $11.95\,\mathrm{ns}$ AFC, the inferred optical depths are $d = 2$ and $d_0 = 0.4$. These values are clearly inconsistent with those seen from the measured spectrum (fig. \ref{combs}(b)), which are around $d = 0.55$ and $d_0 = 0.2$, resulting in a predicted efficiency of around $0.70\%$ (cf. the measured $3.4\,\%$). A similar level of discrepancy is seen for the  $7.97\,\mathrm{ns}$ echo. An explanation that is consistent with this observation is as follows. The AFC here is formed of three narrow velocity distributions in the ground state which results in nine absorption peaks due to the three hyperfine excited state transitions, where in some instances the peaks are overlapping. We therefore have three individual AFCs, one for each transition involved. In the case where the electric fields emitted from these individual combs are in phase they will constructively interfere leading to a 4-fold increase in the detected intensity.

We now consider the limitations of our demonstration. The caesium D2 line has an excited state lifetime of $T_1 = 30.4\,\mathrm{ns}$ \cite{Young1994} representing the ultimate limit on the storage time and the minimum comb tooth spacing $\Delta$. Furthermore, three excited state hyperfine transitions were utilised here thereby placing additional restrictions to integer divisions of the splittings. However, one could operate on the D1 line where the excited-state hyperfine structure is resolved and an arbitrary $\Delta$ could be implemented. The minimal tooth width $\gamma$ we could prepare was $45\,\mathrm{MHz}$ limited by our pumping laser linewidth, pump pulse duration, and power broadening \cite{Main2020} which places a restriction on the minimum $\Delta$, while the Doppler broadening, on the order of $1\,\mathrm{GHz}$, places a restriction on the maximum $\Delta$. So, there is a trade off between operating with a large $\Delta$ to have an AFC rephasing time short with respect to the $T_1$, and operating with a small $\Delta$ to have enough comb teeth to enable a high multimode capacity \cite{Afzelius2009}. 

The peak optical depth here was around $\mathrm{OD}=1$ and a major benefit of our warm vapour platform is that high optical depth is straightforward to achieve \cite{Reim2011} leading to higher light-matter coupling and higher AFC efficiency. An optimised approach to preparing the comb structure could be implemented, for example, adding more pumping modes resonant with the comb in order to ``clean'' the absorbing background \cite{Gundogan2015} thereby increasing the contrast of the comb leading to higher AFC efficiencies. Further, the memory could be placed in a cavity \cite{Saunders2016} and operated in an impedance-matched fashion which can achieve high AFC efficiency \cite{Afzelius2010b}. 

To extend the storage time, one could map the atomic excitation into a long-lived state using a control pulse, in a similar fashion to the full spin-wave AFC protocol \cite{Afzelius2010}. Our implementation lacks the auxiliary ground state to map to, however, there are many suitable upper transitions that could be used in a ladder scheme, similar to the off-resonant cascaded absorption (ORCA) protocol \cite{Kaczmarek2018,Finkelstein2018, Gao2019}. These upper transitions can have relatively strong oscillator strengths allowing $\mathrm{GHz}$ Rabi frequencies to be accessed with moderate control pulse energies. In fact, the AFC is capable of overcoming the inherent Doppler dephasing in these off-resonant Raman-type ladder quantum memories, induced by the large wave-vector mismatch between the signal and control fields. There is potential to go beyond the spin-wave AFC protocol by combining the AFC with Raman/ORCA protocol. After first preparing an AFC, one could map a single photon into coherence between a ground and a doubly-excited state by a two-photon process mediated with an off-resonant control pulse driving the upper transition. The application of a second control pulse at precisely the AFC rephasing time $\tau$ will read out the the memory. If the control pulse is not applied or is mistimed, the memory will not be phase-matched for efficient read-out. The coherence can now be read out at any time multiple of the rephasing time thereby transforming the AFC protocol to an on-demand quantum memory protocol modulo $\tau$ (see \cite{Rubio2018} for a similar proposal and discussion). If a vapour system with a long-lived state (e.g. Rydberg states \cite{Adams2019}) or a metastable atomic level (e.g. barium \cite{Fang2017}) is used as the storage state and the comb is prepared via piecewise adiabatic passage as in \cite{Rubio2018}, the lifetime of the memory will be limited only by the transit time of the atoms across the spatial mode of the beam. For mode sizes around millimetres, several $\mu\mathrm{s}$ could be achieved which is a sufficiently long enough storage time for enhanced quantum operations via temporal multiplexing given the operational bandwidth we have demonstrated \cite{Nunn2013b}.

\begin{acknowledgments} 
In conclusion, we have demonstrated an AFC memory in warm atomic vapour, showing high efficiency  and  multimode  capacity. The flexibility of this atomic system opens the door to a new range of AFC applications and room temperature storage and recall of light in the quantum regime, paving the way for scalable implementations of quantum photonic networking. The authors thank D. Saunders, M. Mazzerra and K. Heshami for useful discussions. We also thank X. Peng and H. Guo for kindly providing the paraffin-coated cell, and R. Zhang for the characterisations of that cell. This work was supported by the UK Engineering and Physical Sciences Research Council (EPSRC) through the Standard Grant No. EP/J000051/1, Programme Grant No. EP/K034480/1, the EPSRC Hub for Networked Quantum Information Technologies (NQIT), and an ERC Advanced Grant (MOQUACINO). TMH is supported via the EPSRC Training and Skills Hub InQuBATE Grant EP/P510270/1.
\end{acknowledgments}

\bibliography{VapeFC}

\begin{thebibliography}{48}%
\makeatletter
\providecommand \@ifxundefined [1]{%
 \@ifx{#1\undefined}
}%
\providecommand \@ifnum [1]{%
 \ifnum #1\expandafter \@firstoftwo
 \else \expandafter \@secondoftwo
 \fi
}%
\providecommand \@ifx [1]{%
 \ifx #1\expandafter \@firstoftwo
 \else \expandafter \@secondoftwo
 \fi
}%
\providecommand \natexlab [1]{#1}%
\providecommand \enquote  [1]{``#1''}%
\providecommand \bibnamefont  [1]{#1}%
\providecommand \bibfnamefont [1]{#1}%
\providecommand \citenamefont [1]{#1}%
\providecommand \href@noop [0]{\@secondoftwo}%
\providecommand \href [0]{\begingroup \@sanitize@url \@href}%
\providecommand \@href[1]{\@@startlink{#1}\@@href}%
\providecommand \@@href[1]{\endgroup#1\@@endlink}%
\providecommand \@sanitize@url [0]{\catcode `\\12\catcode `\$12\catcode
  `\&12\catcode `\#12\catcode `\^12\catcode `\_12\catcode `\%12\relax}%
\providecommand \@@startlink[1]{}%
\providecommand \@@endlink[0]{}%
\providecommand \url  [0]{\begingroup\@sanitize@url \@url }%
\providecommand \@url [1]{\endgroup\@href {#1}{\urlprefix }}%
\providecommand \urlprefix  [0]{URL }%
\providecommand \Eprint [0]{\href }%
\providecommand \doibase [0]{http://dx.doi.org/}%
\providecommand \selectlanguage [0]{\@gobble}%
\providecommand \bibinfo  [0]{\@secondoftwo}%
\providecommand \bibfield  [0]{\@secondoftwo}%
\providecommand \translation [1]{[#1]}%
\providecommand \BibitemOpen [0]{}%
\providecommand \bibitemStop [0]{}%
\providecommand \bibitemNoStop [0]{.\EOS\space}%
\providecommand \EOS [0]{\spacefactor3000\relax}%
\providecommand \BibitemShut  [1]{\csname bibitem#1\endcsname}%
\let\auto@bib@innerbib\@empty
\bibitem [{\citenamefont {Afzelius}\ \emph {et~al.}(2009)\citenamefont
  {Afzelius}, \citenamefont {Simon}, \citenamefont {de~Riedmatten},\ and\
  \citenamefont {Gisin}}]{Afzelius2009}%
  \BibitemOpen
  \bibfield  {author} {\bibinfo {author} {\bibfnamefont {M.}~\bibnamefont
  {Afzelius}}, \bibinfo {author} {\bibfnamefont {C.}~\bibnamefont {Simon}},
  \bibinfo {author} {\bibfnamefont {H.}~\bibnamefont {de~Riedmatten}}, \ and\
  \bibinfo {author} {\bibfnamefont {N.}~\bibnamefont {Gisin}},\ }\href
  {\doibase 10.1103/PhysRevA.79.052329} {\bibfield  {journal} {\bibinfo
  {journal} {Phys. Rev. A}\ }\textbf {\bibinfo {volume} {79}},\ \bibinfo
  {pages} {052329} (\bibinfo {year} {2009})}\BibitemShut {NoStop}%
\bibitem [{\citenamefont {Bussi\`eres}\ \emph {et~al.}(2013)\citenamefont
  {Bussi\`eres}, \citenamefont {Sangouard}, \citenamefont {Afzelius},
  \citenamefont {{de Riedmatten}}, \citenamefont {Simon},\ and\ \citenamefont
  {Tittel}}]{Bussieres2013a}%
  \BibitemOpen
  \bibfield  {author} {\bibinfo {author} {\bibfnamefont {F.}~\bibnamefont
  {Bussi\`eres}}, \bibinfo {author} {\bibfnamefont {N.}~\bibnamefont
  {Sangouard}}, \bibinfo {author} {\bibfnamefont {M.}~\bibnamefont {Afzelius}},
  \bibinfo {author} {\bibfnamefont {H.}~\bibnamefont {{de Riedmatten}}},
  \bibinfo {author} {\bibfnamefont {C.}~\bibnamefont {Simon}}, \ and\ \bibinfo
  {author} {\bibfnamefont {W.}~\bibnamefont {Tittel}},\ }\href {\doibase
  10.1080/09500340.2013.856482} {\bibfield  {journal} {\bibinfo  {journal}
  {Journal of Modern Optics}\ }\textbf {\bibinfo {volume} {60}},\ \bibinfo
  {pages} {1519} (\bibinfo {year} {2013})}\BibitemShut {NoStop}%
\bibitem [{\citenamefont {Heshami}\ \emph {et~al.}(2016)\citenamefont
  {Heshami}, \citenamefont {England}, \citenamefont {Humphreys}, \citenamefont
  {Bustard}, \citenamefont {Acosta}, \citenamefont {Nunn},\ and\ \citenamefont
  {Sussman}}]{Heshami2016}%
  \BibitemOpen
  \bibfield  {author} {\bibinfo {author} {\bibfnamefont {K.}~\bibnamefont
  {Heshami}}, \bibinfo {author} {\bibfnamefont {D.~G.}\ \bibnamefont
  {England}}, \bibinfo {author} {\bibfnamefont {P.~C.}\ \bibnamefont
  {Humphreys}}, \bibinfo {author} {\bibfnamefont {P.~J.}\ \bibnamefont
  {Bustard}}, \bibinfo {author} {\bibfnamefont {V.~M.}\ \bibnamefont {Acosta}},
  \bibinfo {author} {\bibfnamefont {J.}~\bibnamefont {Nunn}}, \ and\ \bibinfo
  {author} {\bibfnamefont {B.~J.}\ \bibnamefont {Sussman}},\ }\href {\doibase
  10.1080/09500340.2016.1148212} {\bibfield  {journal} {\bibinfo  {journal} {J.
  Mod. Opt.}\ }\textbf {\bibinfo {volume} {63}},\ \bibinfo {pages} {2005}
  (\bibinfo {year} {2016})}\BibitemShut {NoStop}%
\bibitem [{\citenamefont {Nunn}\ \emph {et~al.}(2008)\citenamefont {Nunn},
  \citenamefont {Reim}, \citenamefont {Lee}, \citenamefont {Lorenz},
  \citenamefont {Sussman}, \citenamefont {Walmsley},\ and\ \citenamefont
  {Jaksch}}]{Nunn2008}%
  \BibitemOpen
  \bibfield  {author} {\bibinfo {author} {\bibfnamefont {J.}~\bibnamefont
  {Nunn}}, \bibinfo {author} {\bibfnamefont {K.}~\bibnamefont {Reim}}, \bibinfo
  {author} {\bibfnamefont {K.~C.}\ \bibnamefont {Lee}}, \bibinfo {author}
  {\bibfnamefont {V.~O.}\ \bibnamefont {Lorenz}}, \bibinfo {author}
  {\bibfnamefont {B.~J.}\ \bibnamefont {Sussman}}, \bibinfo {author}
  {\bibfnamefont {I.~A.}\ \bibnamefont {Walmsley}}, \ and\ \bibinfo {author}
  {\bibfnamefont {D.}~\bibnamefont {Jaksch}},\ }\href {\doibase
  10.1103/PhysRevLett.101.260502} {\bibfield  {journal} {\bibinfo  {journal}
  {Phys. Rev. Lett.}\ }\textbf {\bibinfo {volume} {101}},\ \bibinfo {pages}
  {260502} (\bibinfo {year} {2008})}\BibitemShut {NoStop}%
\bibitem [{\citenamefont {Simon}\ \emph {et~al.}(2007)\citenamefont {Simon},
  \citenamefont {de~Riedmatten}, \citenamefont {Afzelius}, \citenamefont
  {Sangouard}, \citenamefont {Zbinden},\ and\ \citenamefont
  {Gisin}}]{Simon2007}%
  \BibitemOpen
  \bibfield  {author} {\bibinfo {author} {\bibfnamefont {C.}~\bibnamefont
  {Simon}}, \bibinfo {author} {\bibfnamefont {H.}~\bibnamefont
  {de~Riedmatten}}, \bibinfo {author} {\bibfnamefont {M.}~\bibnamefont
  {Afzelius}}, \bibinfo {author} {\bibfnamefont {N.}~\bibnamefont {Sangouard}},
  \bibinfo {author} {\bibfnamefont {H.}~\bibnamefont {Zbinden}}, \ and\
  \bibinfo {author} {\bibfnamefont {N.}~\bibnamefont {Gisin}},\ }\href
  {\doibase 10.1103/PhysRevLett.98.190503} {\bibfield  {journal} {\bibinfo
  {journal} {Phys. Rev. Lett.}\ }\textbf {\bibinfo {volume} {98}},\ \bibinfo
  {pages} {190503} (\bibinfo {year} {2007})}\BibitemShut {NoStop}%
\bibitem [{\citenamefont {G\"undo{\u g}an}\ \emph {et~al.}(2012)\citenamefont
  {G\"undo{\u g}an}, \citenamefont {Ledingham}, \citenamefont {Almasi},
  \citenamefont {Cristiani},\ and\ \citenamefont {{de
  Riedmatten}}}]{Gundogan2012}%
  \BibitemOpen
  \bibfield  {author} {\bibinfo {author} {\bibfnamefont {M.}~\bibnamefont
  {G\"undo{\u g}an}}, \bibinfo {author} {\bibfnamefont {P.~M.}\ \bibnamefont
  {Ledingham}}, \bibinfo {author} {\bibfnamefont {A.}~\bibnamefont {Almasi}},
  \bibinfo {author} {\bibfnamefont {M.}~\bibnamefont {Cristiani}}, \ and\
  \bibinfo {author} {\bibfnamefont {H.}~\bibnamefont {{de Riedmatten}}},\
  }\href {\doibase 10.1103/PhysRevLett.108.190504} {\bibfield  {journal}
  {\bibinfo  {journal} {Phys. Rev. Lett.}\ }\textbf {\bibinfo {volume} {108}},\
  \bibinfo {pages} {190504} (\bibinfo {year} {2012})}\BibitemShut {NoStop}%
\bibitem [{\citenamefont {Zhou}\ \emph {et~al.}(2012)\citenamefont {Zhou},
  \citenamefont {Lin}, \citenamefont {Yang}, \citenamefont {Li},\ and\
  \citenamefont {Guo}}]{Zhou2012}%
  \BibitemOpen
  \bibfield  {author} {\bibinfo {author} {\bibfnamefont {Z.-Q.}\ \bibnamefont
  {Zhou}}, \bibinfo {author} {\bibfnamefont {W.-B.}\ \bibnamefont {Lin}},
  \bibinfo {author} {\bibfnamefont {M.}~\bibnamefont {Yang}}, \bibinfo {author}
  {\bibfnamefont {C.-F.}\ \bibnamefont {Li}}, \ and\ \bibinfo {author}
  {\bibfnamefont {G.-C.}\ \bibnamefont {Guo}},\ }\href {\doibase
  10.1103/PhysRevLett.108.190505} {\bibfield  {journal} {\bibinfo  {journal}
  {Phys. Rev. Lett.}\ }\textbf {\bibinfo {volume} {108}},\ \bibinfo {pages}
  {190505} (\bibinfo {year} {2012})}\BibitemShut {NoStop}%
\bibitem [{\citenamefont {Clausen}\ \emph {et~al.}(2012)\citenamefont
  {Clausen}, \citenamefont {Bussi\`eres}, \citenamefont {Afzelius},\ and\
  \citenamefont {Gisin}}]{Clausen2012}%
  \BibitemOpen
  \bibfield  {author} {\bibinfo {author} {\bibfnamefont {C.}~\bibnamefont
  {Clausen}}, \bibinfo {author} {\bibfnamefont {F.}~\bibnamefont
  {Bussi\`eres}}, \bibinfo {author} {\bibfnamefont {M.}~\bibnamefont
  {Afzelius}}, \ and\ \bibinfo {author} {\bibfnamefont {N.}~\bibnamefont
  {Gisin}},\ }\href {\doibase 10.1103/PhysRevLett.108.190503} {\bibfield
  {journal} {\bibinfo  {journal} {Phys. Rev. Lett.}\ }\textbf {\bibinfo
  {volume} {108}},\ \bibinfo {pages} {190503} (\bibinfo {year}
  {2012})}\BibitemShut {NoStop}%
\bibitem [{\citenamefont {G\"und{\v o}gan}\ \emph {et~al.}(2015)\citenamefont
  {G\"und{\v o}gan}, \citenamefont {Ledingham}, \citenamefont {Kutluer},
  \citenamefont {Mazzera},\ and\ \citenamefont {De~Riedmatten}}]{Gundogan2015}%
  \BibitemOpen
  \bibfield  {author} {\bibinfo {author} {\bibfnamefont {M.}~\bibnamefont
  {G\"und{\v o}gan}}, \bibinfo {author} {\bibfnamefont {P.~M.}\ \bibnamefont
  {Ledingham}}, \bibinfo {author} {\bibfnamefont {K.}~\bibnamefont {Kutluer}},
  \bibinfo {author} {\bibfnamefont {M.}~\bibnamefont {Mazzera}}, \ and\
  \bibinfo {author} {\bibfnamefont {H.}~\bibnamefont {De~Riedmatten}},\ }\href
  {\doibase 10.1103/PhysRevLett.114.230501} {\bibfield  {journal} {\bibinfo
  {journal} {Phys. Rev. Lett.}\ }\textbf {\bibinfo {volume} {114}},\ \bibinfo
  {pages} {1} (\bibinfo {year} {2015})}\BibitemShut {NoStop}%
\bibitem [{\citenamefont {Hua}\ \emph {et~al.}(2019)\citenamefont {Hua},
  \citenamefont {Yang}, \citenamefont {Zhou}, \citenamefont {Wang},
  \citenamefont {Liu}, \citenamefont {Li}, \citenamefont {Li}, \citenamefont
  {Ma}, \citenamefont {Liu}, \citenamefont {Liang}, \citenamefont {Hu},
  \citenamefont {Li}, \citenamefont {Li},\ and\ \citenamefont {Guo}}]{Hua2019}%
  \BibitemOpen
  \bibfield  {author} {\bibinfo {author} {\bibfnamefont {Y.-L.}\ \bibnamefont
  {Hua}}, \bibinfo {author} {\bibfnamefont {T.-S.}\ \bibnamefont {Yang}},
  \bibinfo {author} {\bibfnamefont {Z.-Q.}\ \bibnamefont {Zhou}}, \bibinfo
  {author} {\bibfnamefont {J.}~\bibnamefont {Wang}}, \bibinfo {author}
  {\bibfnamefont {X.}~\bibnamefont {Liu}}, \bibinfo {author} {\bibfnamefont
  {Z.-F.}\ \bibnamefont {Li}}, \bibinfo {author} {\bibfnamefont {P.-Y.}\
  \bibnamefont {Li}}, \bibinfo {author} {\bibfnamefont {Y.}~\bibnamefont {Ma}},
  \bibinfo {author} {\bibfnamefont {C.}~\bibnamefont {Liu}}, \bibinfo {author}
  {\bibfnamefont {P.-J.}\ \bibnamefont {Liang}}, \bibinfo {author}
  {\bibfnamefont {J.}~\bibnamefont {Hu}}, \bibinfo {author} {\bibfnamefont
  {X.}~\bibnamefont {Li}}, \bibinfo {author} {\bibfnamefont {C.-F.}\
  \bibnamefont {Li}}, \ and\ \bibinfo {author} {\bibfnamefont {G.-C.}\
  \bibnamefont {Guo}},\ }\href {\doibase
  https://doi.org/10.1016/j.scib.2019.09.006} {\bibfield  {journal} {\bibinfo
  {journal} {Science Bulletin}\ }\textbf {\bibinfo {volume} {64}},\ \bibinfo
  {pages} {1577 } (\bibinfo {year} {2019})}\BibitemShut {NoStop}%
\bibitem [{\citenamefont {Saglamyurek}\ \emph {et~al.}(2011)\citenamefont
  {Saglamyurek}, \citenamefont {Sinclair}, \citenamefont {Jin}, \citenamefont
  {Slater}, \citenamefont {Oblak}, \citenamefont {Bussi\`eres}, \citenamefont
  {George}, \citenamefont {Ricken}, \citenamefont {Sohler},\ and\ \citenamefont
  {Tittel}}]{Saglamyurek2011}%
  \BibitemOpen
  \bibfield  {author} {\bibinfo {author} {\bibfnamefont {E.}~\bibnamefont
  {Saglamyurek}}, \bibinfo {author} {\bibfnamefont {N.}~\bibnamefont
  {Sinclair}}, \bibinfo {author} {\bibfnamefont {J.}~\bibnamefont {Jin}},
  \bibinfo {author} {\bibfnamefont {J.~A.}\ \bibnamefont {Slater}}, \bibinfo
  {author} {\bibfnamefont {D.}~\bibnamefont {Oblak}}, \bibinfo {author}
  {\bibfnamefont {F.}~\bibnamefont {Bussi\`eres}}, \bibinfo {author}
  {\bibfnamefont {M.}~\bibnamefont {George}}, \bibinfo {author} {\bibfnamefont
  {R.}~\bibnamefont {Ricken}}, \bibinfo {author} {\bibfnamefont
  {W.}~\bibnamefont {Sohler}}, \ and\ \bibinfo {author} {\bibfnamefont
  {W.}~\bibnamefont {Tittel}},\ }\href {\doibase 10.1038/nature09719}
  {\bibfield  {journal} {\bibinfo  {journal} {Nature}\ }\textbf {\bibinfo
  {volume} {469}},\ \bibinfo {pages} {512} (\bibinfo {year}
  {2011})}\BibitemShut {NoStop}%
\bibitem [{\citenamefont {de~Riedmatten}\ \emph {et~al.}(2008)\citenamefont
  {de~Riedmatten}, \citenamefont {Afzelius}, \citenamefont {Staudt},
  \citenamefont {Simon},\ and\ \citenamefont {Gisin}}]{DeRiedmatten2008}%
  \BibitemOpen
  \bibfield  {author} {\bibinfo {author} {\bibfnamefont {H.}~\bibnamefont
  {de~Riedmatten}}, \bibinfo {author} {\bibfnamefont {M.}~\bibnamefont
  {Afzelius}}, \bibinfo {author} {\bibfnamefont {M.~U.}\ \bibnamefont
  {Staudt}}, \bibinfo {author} {\bibfnamefont {C.}~\bibnamefont {Simon}}, \
  and\ \bibinfo {author} {\bibfnamefont {N.}~\bibnamefont {Gisin}},\ }\href
  {\doibase 10.1038/nature07607} {\bibfield  {journal} {\bibinfo  {journal}
  {Nature}\ }\textbf {\bibinfo {volume} {456}},\ \bibinfo {pages} {773}
  (\bibinfo {year} {2008})}\BibitemShut {NoStop}%
\bibitem [{\citenamefont {Riel\"ander}\ \emph {et~al.}(2014)\citenamefont
  {Riel\"ander}, \citenamefont {Kutluer}, \citenamefont {Ledingham},
  \citenamefont {G\"undo{\u g}an}, \citenamefont {Fekete}, \citenamefont
  {Mazzera},\ and\ \citenamefont {{de Riedmatten}}}]{Rielander2014}%
  \BibitemOpen
  \bibfield  {author} {\bibinfo {author} {\bibfnamefont {D.}~\bibnamefont
  {Riel\"ander}}, \bibinfo {author} {\bibfnamefont {K.}~\bibnamefont
  {Kutluer}}, \bibinfo {author} {\bibfnamefont {P.~M.}\ \bibnamefont
  {Ledingham}}, \bibinfo {author} {\bibfnamefont {M.}~\bibnamefont {G\"undo{\u
  g}an}}, \bibinfo {author} {\bibfnamefont {J.}~\bibnamefont {Fekete}},
  \bibinfo {author} {\bibfnamefont {M.}~\bibnamefont {Mazzera}}, \ and\
  \bibinfo {author} {\bibfnamefont {H.}~\bibnamefont {{de Riedmatten}}},\
  }\href {\doibase 10.1103/PhysRevLett.112.040504} {\bibfield  {journal}
  {\bibinfo  {journal} {Phys. Rev. Lett.}\ }\textbf {\bibinfo {volume} {112}},\
  \bibinfo {pages} {040504} (\bibinfo {year} {2014})}\BibitemShut {NoStop}%
\bibitem [{\citenamefont {Tang}\ \emph {et~al.}(2015)\citenamefont {Tang},
  \citenamefont {Zhou}, \citenamefont {Wang}, \citenamefont {Li}, \citenamefont
  {Liu}, \citenamefont {Hua}, \citenamefont {Zou}, \citenamefont {Wang},
  \citenamefont {He}, \citenamefont {Chen}, \citenamefont {Sun}, \citenamefont
  {Yu}, \citenamefont {Li}, \citenamefont {Zha}, \citenamefont {Ni},
  \citenamefont {Niu}, \citenamefont {Li},\ and\ \citenamefont
  {Guo}}]{Tang2015}%
  \BibitemOpen
  \bibfield  {author} {\bibinfo {author} {\bibfnamefont {J.-S.}\ \bibnamefont
  {Tang}}, \bibinfo {author} {\bibfnamefont {Z.-Q.}\ \bibnamefont {Zhou}},
  \bibinfo {author} {\bibfnamefont {Y.-T.}\ \bibnamefont {Wang}}, \bibinfo
  {author} {\bibfnamefont {Y.-L.}\ \bibnamefont {Li}}, \bibinfo {author}
  {\bibfnamefont {X.}~\bibnamefont {Liu}}, \bibinfo {author} {\bibfnamefont
  {Y.-L.}\ \bibnamefont {Hua}}, \bibinfo {author} {\bibfnamefont
  {Y.}~\bibnamefont {Zou}}, \bibinfo {author} {\bibfnamefont {S.}~\bibnamefont
  {Wang}}, \bibinfo {author} {\bibfnamefont {D.-Y.}\ \bibnamefont {He}},
  \bibinfo {author} {\bibfnamefont {G.}~\bibnamefont {Chen}}, \bibinfo {author}
  {\bibfnamefont {Y.-N.}\ \bibnamefont {Sun}}, \bibinfo {author} {\bibfnamefont
  {Y.}~\bibnamefont {Yu}}, \bibinfo {author} {\bibfnamefont {M.-F.}\
  \bibnamefont {Li}}, \bibinfo {author} {\bibfnamefont {G.-W.}\ \bibnamefont
  {Zha}}, \bibinfo {author} {\bibfnamefont {H.-Q.}\ \bibnamefont {Ni}},
  \bibinfo {author} {\bibfnamefont {Z.-C.}\ \bibnamefont {Niu}}, \bibinfo
  {author} {\bibfnamefont {C.-F.}\ \bibnamefont {Li}}, \ and\ \bibinfo {author}
  {\bibfnamefont {G.-C.}\ \bibnamefont {Guo}},\ }\href {\doibase
  10.1038/ncomms9652} {\bibfield  {journal} {\bibinfo  {journal} {Nature
  Communications}\ }\textbf {\bibinfo {volume} {6}},\ \bibinfo {pages} {8652}
  (\bibinfo {year} {2015})}\BibitemShut {NoStop}%
\bibitem [{\citenamefont {Tiranov}\ \emph {et~al.}(2016)\citenamefont
  {Tiranov}, \citenamefont {Strassmann}, \citenamefont {Lavoie}, \citenamefont
  {Brunner}, \citenamefont {Huber}, \citenamefont {Verma}, \citenamefont {Nam},
  \citenamefont {Mirin}, \citenamefont {Lita}, \citenamefont {Marsili},
  \citenamefont {Afzelius}, \citenamefont {Bussi\`eres},\ and\ \citenamefont
  {Gisin}}]{Tiranov2016}%
  \BibitemOpen
  \bibfield  {author} {\bibinfo {author} {\bibfnamefont {A.}~\bibnamefont
  {Tiranov}}, \bibinfo {author} {\bibfnamefont {P.~C.}\ \bibnamefont
  {Strassmann}}, \bibinfo {author} {\bibfnamefont {J.}~\bibnamefont {Lavoie}},
  \bibinfo {author} {\bibfnamefont {N.}~\bibnamefont {Brunner}}, \bibinfo
  {author} {\bibfnamefont {M.}~\bibnamefont {Huber}}, \bibinfo {author}
  {\bibfnamefont {V.~B.}\ \bibnamefont {Verma}}, \bibinfo {author}
  {\bibfnamefont {S.~W.}\ \bibnamefont {Nam}}, \bibinfo {author} {\bibfnamefont
  {R.~P.}\ \bibnamefont {Mirin}}, \bibinfo {author} {\bibfnamefont {A.~E.}\
  \bibnamefont {Lita}}, \bibinfo {author} {\bibfnamefont {F.}~\bibnamefont
  {Marsili}}, \bibinfo {author} {\bibfnamefont {M.}~\bibnamefont {Afzelius}},
  \bibinfo {author} {\bibfnamefont {F.}~\bibnamefont {Bussi\`eres}}, \ and\
  \bibinfo {author} {\bibfnamefont {N.}~\bibnamefont {Gisin}},\ }\href
  {\doibase 10.1103/PhysRevLett.117.240506} {\bibfield  {journal} {\bibinfo
  {journal} {Phys. Rev. Lett.}\ }\textbf {\bibinfo {volume} {117}},\ \bibinfo
  {pages} {240506} (\bibinfo {year} {2016})}\BibitemShut {NoStop}%
\bibitem [{\citenamefont {Tiranov}\ \emph {et~al.}(2015)\citenamefont
  {Tiranov}, \citenamefont {Lavoie}, \citenamefont {Ferrier}, \citenamefont
  {Goldner}, \citenamefont {Verma}, \citenamefont {Nam}, \citenamefont {Mirin},
  \citenamefont {Lita}, \citenamefont {Marsili}, \citenamefont {Herrmann},
  \citenamefont {Silberhorn}, \citenamefont {Gisin}, \citenamefont {Afzelius},\
  and\ \citenamefont {Bussi\`{e}res}}]{Tiranov15}%
  \BibitemOpen
  \bibfield  {author} {\bibinfo {author} {\bibfnamefont {A.}~\bibnamefont
  {Tiranov}}, \bibinfo {author} {\bibfnamefont {J.}~\bibnamefont {Lavoie}},
  \bibinfo {author} {\bibfnamefont {A.}~\bibnamefont {Ferrier}}, \bibinfo
  {author} {\bibfnamefont {P.}~\bibnamefont {Goldner}}, \bibinfo {author}
  {\bibfnamefont {V.~B.}\ \bibnamefont {Verma}}, \bibinfo {author}
  {\bibfnamefont {S.~W.}\ \bibnamefont {Nam}}, \bibinfo {author} {\bibfnamefont
  {R.~P.}\ \bibnamefont {Mirin}}, \bibinfo {author} {\bibfnamefont {A.~E.}\
  \bibnamefont {Lita}}, \bibinfo {author} {\bibfnamefont {F.}~\bibnamefont
  {Marsili}}, \bibinfo {author} {\bibfnamefont {H.}~\bibnamefont {Herrmann}},
  \bibinfo {author} {\bibfnamefont {C.}~\bibnamefont {Silberhorn}}, \bibinfo
  {author} {\bibfnamefont {N.}~\bibnamefont {Gisin}}, \bibinfo {author}
  {\bibfnamefont {M.}~\bibnamefont {Afzelius}}, \ and\ \bibinfo {author}
  {\bibfnamefont {F.}~\bibnamefont {Bussi\`{e}res}},\ }\href {\doibase
  10.1364/OPTICA.2.000279} {\bibfield  {journal} {\bibinfo  {journal} {Optica}\
  }\textbf {\bibinfo {volume} {2}},\ \bibinfo {pages} {279} (\bibinfo {year}
  {2015})}\BibitemShut {NoStop}%
\bibitem [{\citenamefont {Kutluer}\ \emph {et~al.}(2017)\citenamefont
  {Kutluer}, \citenamefont {Mazzera},\ and\ \citenamefont
  {de~Riedmatten}}]{Kutluer2017}%
  \BibitemOpen
  \bibfield  {author} {\bibinfo {author} {\bibfnamefont {K.}~\bibnamefont
  {Kutluer}}, \bibinfo {author} {\bibfnamefont {M.}~\bibnamefont {Mazzera}}, \
  and\ \bibinfo {author} {\bibfnamefont {H.}~\bibnamefont {de~Riedmatten}},\
  }\href {\doibase 10.1103/PhysRevLett.118.210502} {\bibfield  {journal}
  {\bibinfo  {journal} {Phys. Rev. Lett.}\ }\textbf {\bibinfo {volume} {118}},\
  \bibinfo {pages} {210502} (\bibinfo {year} {2017})}\BibitemShut {NoStop}%
\bibitem [{\citenamefont {Laplane}\ \emph {et~al.}(2017)\citenamefont
  {Laplane}, \citenamefont {Jobez}, \citenamefont {Etesse}, \citenamefont
  {Gisin},\ and\ \citenamefont {Afzelius}}]{Laplane2017}%
  \BibitemOpen
  \bibfield  {author} {\bibinfo {author} {\bibfnamefont {C.}~\bibnamefont
  {Laplane}}, \bibinfo {author} {\bibfnamefont {P.}~\bibnamefont {Jobez}},
  \bibinfo {author} {\bibfnamefont {J.}~\bibnamefont {Etesse}}, \bibinfo
  {author} {\bibfnamefont {N.}~\bibnamefont {Gisin}}, \ and\ \bibinfo {author}
  {\bibfnamefont {M.}~\bibnamefont {Afzelius}},\ }\href {\doibase
  10.1103/PhysRevLett.118.210501} {\bibfield  {journal} {\bibinfo  {journal}
  {Phys. Rev. Lett.}\ }\textbf {\bibinfo {volume} {118}},\ \bibinfo {pages}
  {210501} (\bibinfo {year} {2017})}\BibitemShut {NoStop}%
\bibitem [{\citenamefont {Usmani}\ \emph {et~al.}(2010)\citenamefont {Usmani},
  \citenamefont {Afzelius}, \citenamefont {de~Riedmatten},\ and\ \citenamefont
  {Gisin}}]{Usmani2010}%
  \BibitemOpen
  \bibfield  {author} {\bibinfo {author} {\bibfnamefont {I.}~\bibnamefont
  {Usmani}}, \bibinfo {author} {\bibfnamefont {M.}~\bibnamefont {Afzelius}},
  \bibinfo {author} {\bibfnamefont {H.}~\bibnamefont {de~Riedmatten}}, \ and\
  \bibinfo {author} {\bibfnamefont {N.}~\bibnamefont {Gisin}},\ }\href
  {\doibase 10.1038/ncomms1010} {\bibfield  {journal} {\bibinfo  {journal}
  {Nature Communications}\ }\textbf {\bibinfo {volume} {1}},\ \bibinfo {pages}
  {12} (\bibinfo {year} {2010})}\BibitemShut {NoStop}%
\bibitem [{\citenamefont {Seri}\ \emph {et~al.}(2019)\citenamefont {Seri},
  \citenamefont {Lago-Rivera}, \citenamefont {Lenhard}, \citenamefont
  {Corrielli}, \citenamefont {Osellame}, \citenamefont {Mazzera},\ and\
  \citenamefont {de~Riedmatten}}]{Seri2019}%
  \BibitemOpen
  \bibfield  {author} {\bibinfo {author} {\bibfnamefont {A.}~\bibnamefont
  {Seri}}, \bibinfo {author} {\bibfnamefont {D.}~\bibnamefont {Lago-Rivera}},
  \bibinfo {author} {\bibfnamefont {A.}~\bibnamefont {Lenhard}}, \bibinfo
  {author} {\bibfnamefont {G.}~\bibnamefont {Corrielli}}, \bibinfo {author}
  {\bibfnamefont {R.}~\bibnamefont {Osellame}}, \bibinfo {author}
  {\bibfnamefont {M.}~\bibnamefont {Mazzera}}, \ and\ \bibinfo {author}
  {\bibfnamefont {H.}~\bibnamefont {de~Riedmatten}},\ }\href {\doibase
  10.1103/PhysRevLett.123.080502} {\bibfield  {journal} {\bibinfo  {journal}
  {Phys. Rev. Lett.}\ }\textbf {\bibinfo {volume} {123}},\ \bibinfo {pages}
  {080502} (\bibinfo {year} {2019})}\BibitemShut {NoStop}%
\bibitem [{\citenamefont {McAuslan}\ \emph {et~al.}(2012)\citenamefont
  {McAuslan}, \citenamefont {Taylor},\ and\ \citenamefont
  {Longdell}}]{McAuslan2012}%
  \BibitemOpen
  \bibfield  {author} {\bibinfo {author} {\bibfnamefont {D.~L.}\ \bibnamefont
  {McAuslan}}, \bibinfo {author} {\bibfnamefont {L.~R.}\ \bibnamefont
  {Taylor}}, \ and\ \bibinfo {author} {\bibfnamefont {J.~J.}\ \bibnamefont
  {Longdell}},\ }\href {\doibase 10.1063/1.4766341} {\bibfield  {journal}
  {\bibinfo  {journal} {Applied Physics Letters}\ }\textbf {\bibinfo {volume}
  {101}},\ \bibinfo {pages} {191112} (\bibinfo {year} {2012})}\BibitemShut
  {NoStop}%
\bibitem [{\citenamefont {Thomas}\ \emph {et~al.}(2017)\citenamefont {Thomas},
  \citenamefont {Munns}, \citenamefont {Kaczmarek}, \citenamefont {Qiu},
  \citenamefont {Brecht}, \citenamefont {Feizpour}, \citenamefont {Ledingham},
  \citenamefont {Walmsley}, \citenamefont {Nunn},\ and\ \citenamefont
  {Saunders}}]{Thomas2017}%
  \BibitemOpen
  \bibfield  {author} {\bibinfo {author} {\bibfnamefont {S.~E.}\ \bibnamefont
  {Thomas}}, \bibinfo {author} {\bibfnamefont {J.~H.~D.}\ \bibnamefont
  {Munns}}, \bibinfo {author} {\bibfnamefont {K.~T.}\ \bibnamefont
  {Kaczmarek}}, \bibinfo {author} {\bibfnamefont {C.}~\bibnamefont {Qiu}},
  \bibinfo {author} {\bibfnamefont {B.}~\bibnamefont {Brecht}}, \bibinfo
  {author} {\bibfnamefont {A.}~\bibnamefont {Feizpour}}, \bibinfo {author}
  {\bibfnamefont {P.~M.}\ \bibnamefont {Ledingham}}, \bibinfo {author}
  {\bibfnamefont {I.~A.}\ \bibnamefont {Walmsley}}, \bibinfo {author}
  {\bibfnamefont {J.}~\bibnamefont {Nunn}}, \ and\ \bibinfo {author}
  {\bibfnamefont {D.~J.}\ \bibnamefont {Saunders}},\ }\href
  {http://stacks.iop.org/1367-2630/19/i=6/a=063034} {\bibfield  {journal}
  {\bibinfo  {journal} {New J. Phys.}\ }\textbf {\bibinfo {volume} {19}},\
  \bibinfo {pages} {063034} (\bibinfo {year} {2017})}\BibitemShut {NoStop}%
\bibitem [{\citenamefont {Wolters}\ \emph {et~al.}(2017)\citenamefont
  {Wolters}, \citenamefont {Buser}, \citenamefont {Horsley}, \citenamefont
  {B\'eguin}, \citenamefont {J\"ockel}, \citenamefont {Jahn}, \citenamefont
  {Warburton},\ and\ \citenamefont {Treutlein}}]{Wolters2017}%
  \BibitemOpen
  \bibfield  {author} {\bibinfo {author} {\bibfnamefont {J.}~\bibnamefont
  {Wolters}}, \bibinfo {author} {\bibfnamefont {G.}~\bibnamefont {Buser}},
  \bibinfo {author} {\bibfnamefont {A.}~\bibnamefont {Horsley}}, \bibinfo
  {author} {\bibfnamefont {L.}~\bibnamefont {B\'eguin}}, \bibinfo {author}
  {\bibfnamefont {A.}~\bibnamefont {J\"ockel}}, \bibinfo {author}
  {\bibfnamefont {J.-P.}\ \bibnamefont {Jahn}}, \bibinfo {author}
  {\bibfnamefont {R.~J.}\ \bibnamefont {Warburton}}, \ and\ \bibinfo {author}
  {\bibfnamefont {P.}~\bibnamefont {Treutlein}},\ }\href {\doibase
  10.1103/PhysRevLett.119.060502} {\bibfield  {journal} {\bibinfo  {journal}
  {Phys. Rev. Lett.}\ }\textbf {\bibinfo {volume} {119}},\ \bibinfo {pages}
  {060502} (\bibinfo {year} {2017})}\BibitemShut {NoStop}%
\bibitem [{\citenamefont {Namazi}\ \emph {et~al.}(2017)\citenamefont {Namazi},
  \citenamefont {Kupchak}, \citenamefont {Jordaan}, \citenamefont
  {Shahrokhshahi},\ and\ \citenamefont {Figueroa}}]{Namazi2017}%
  \BibitemOpen
  \bibfield  {author} {\bibinfo {author} {\bibfnamefont {M.}~\bibnamefont
  {Namazi}}, \bibinfo {author} {\bibfnamefont {C.}~\bibnamefont {Kupchak}},
  \bibinfo {author} {\bibfnamefont {B.}~\bibnamefont {Jordaan}}, \bibinfo
  {author} {\bibfnamefont {R.}~\bibnamefont {Shahrokhshahi}}, \ and\ \bibinfo
  {author} {\bibfnamefont {E.}~\bibnamefont {Figueroa}},\ }\href {\doibase
  10.1103/PhysRevApplied.8.034023} {\bibfield  {journal} {\bibinfo  {journal}
  {Phys. Rev. Applied}\ }\textbf {\bibinfo {volume} {8}},\ \bibinfo {pages}
  {034023} (\bibinfo {year} {2017})}\BibitemShut {NoStop}%
\bibitem [{\citenamefont {Hosseini}\ \emph {et~al.}(2011)\citenamefont
  {Hosseini}, \citenamefont {Sparkes}, \citenamefont {Campbell}, \citenamefont
  {Lam},\ and\ \citenamefont {Buchler}}]{Hosseini2011}%
  \BibitemOpen
  \bibfield  {author} {\bibinfo {author} {\bibfnamefont {M.}~\bibnamefont
  {Hosseini}}, \bibinfo {author} {\bibfnamefont {B.~M.}\ \bibnamefont
  {Sparkes}}, \bibinfo {author} {\bibfnamefont {G.}~\bibnamefont {Campbell}},
  \bibinfo {author} {\bibfnamefont {P.~K.}\ \bibnamefont {Lam}}, \ and\
  \bibinfo {author} {\bibfnamefont {B.~C.}\ \bibnamefont {Buchler}},\ }\href
  {\doibase 10.1038/ncomms1175} {\bibfield  {journal} {\bibinfo  {journal}
  {Nature Communications}\ }\textbf {\bibinfo {volume} {2}},\ \bibinfo {pages}
  {174} (\bibinfo {year} {2011})}\BibitemShut {NoStop}%
\bibitem [{\citenamefont {Reim}\ \emph {et~al.}(2011)\citenamefont {Reim},
  \citenamefont {Michelberger}, \citenamefont {Lee}, \citenamefont {Nunn},
  \citenamefont {Langford},\ and\ \citenamefont {Walmsley}}]{Reim2011}%
  \BibitemOpen
  \bibfield  {author} {\bibinfo {author} {\bibfnamefont {K.~F.}\ \bibnamefont
  {Reim}}, \bibinfo {author} {\bibfnamefont {P.}~\bibnamefont {Michelberger}},
  \bibinfo {author} {\bibfnamefont {K.~C.}\ \bibnamefont {Lee}}, \bibinfo
  {author} {\bibfnamefont {J.}~\bibnamefont {Nunn}}, \bibinfo {author}
  {\bibfnamefont {N.~K.}\ \bibnamefont {Langford}}, \ and\ \bibinfo {author}
  {\bibfnamefont {I.~A.}\ \bibnamefont {Walmsley}},\ }\href {\doibase
  10.1103/PhysRevLett.107.053603} {\bibfield  {journal} {\bibinfo  {journal}
  {Phys. Rev. Lett.}\ }\textbf {\bibinfo {volume} {107}},\ \bibinfo {pages}
  {053603} (\bibinfo {year} {2011})}\BibitemShut {NoStop}%
\bibitem [{\citenamefont {Thomas}\ \emph {et~al.}(2019)\citenamefont {Thomas},
  \citenamefont {Hird}, \citenamefont {Munns}, \citenamefont {Brecht},
  \citenamefont {Saunders}, \citenamefont {Nunn}, \citenamefont {Walmsley},\
  and\ \citenamefont {Ledingham}}]{Thomas2019}%
  \BibitemOpen
  \bibfield  {author} {\bibinfo {author} {\bibfnamefont {S.~E.}\ \bibnamefont
  {Thomas}}, \bibinfo {author} {\bibfnamefont {T.~M.}\ \bibnamefont {Hird}},
  \bibinfo {author} {\bibfnamefont {J.~H.~D.}\ \bibnamefont {Munns}}, \bibinfo
  {author} {\bibfnamefont {B.}~\bibnamefont {Brecht}}, \bibinfo {author}
  {\bibfnamefont {D.~J.}\ \bibnamefont {Saunders}}, \bibinfo {author}
  {\bibfnamefont {J.}~\bibnamefont {Nunn}}, \bibinfo {author} {\bibfnamefont
  {I.~A.}\ \bibnamefont {Walmsley}}, \ and\ \bibinfo {author} {\bibfnamefont
  {P.~M.}\ \bibnamefont {Ledingham}},\ }\href {\doibase
  10.1103/PhysRevA.100.033801} {\bibfield  {journal} {\bibinfo  {journal}
  {Phys. Rev. A}\ }\textbf {\bibinfo {volume} {100}},\ \bibinfo {pages}
  {033801} (\bibinfo {year} {2019})}\BibitemShut {NoStop}%
\bibitem [{\citenamefont {Preston}(1996)}]{Preston1996}%
  \BibitemOpen
  \bibfield  {author} {\bibinfo {author} {\bibfnamefont {D.~W.}\ \bibnamefont
  {Preston}},\ }\href {\doibase 10.1119/1.18457} {\bibfield  {journal}
  {\bibinfo  {journal} {American Journal of Physics}\ }\textbf {\bibinfo
  {volume} {64}},\ \bibinfo {pages} {1432} (\bibinfo {year}
  {1996})}\BibitemShut {NoStop}%
\bibitem [{\citenamefont {Marian}\ \emph {et~al.}(2004)\citenamefont {Marian},
  \citenamefont {Stowe}, \citenamefont {Lawall}, \citenamefont {Felinto},\ and\
  \citenamefont {Ye}}]{Marian2004}%
  \BibitemOpen
  \bibfield  {author} {\bibinfo {author} {\bibfnamefont {A.}~\bibnamefont
  {Marian}}, \bibinfo {author} {\bibfnamefont {M.~C.}\ \bibnamefont {Stowe}},
  \bibinfo {author} {\bibfnamefont {J.~R.}\ \bibnamefont {Lawall}}, \bibinfo
  {author} {\bibfnamefont {D.}~\bibnamefont {Felinto}}, \ and\ \bibinfo
  {author} {\bibfnamefont {J.}~\bibnamefont {Ye}},\ }\href {\doibase
  10.1126/science.1105660} {\bibfield  {journal} {\bibinfo  {journal}
  {Science}\ }\textbf {\bibinfo {volume} {306}},\ \bibinfo {pages} {2063}
  (\bibinfo {year} {2004})}\BibitemShut {NoStop}%
\bibitem [{\citenamefont {Aumiler}\ \emph {et~al.}(2005)\citenamefont
  {Aumiler}, \citenamefont {Ban}, \citenamefont
  {Skenderovi\ifmmode~\acute{c}\else \'{c}\fi{}},\ and\ \citenamefont
  {Pichler}}]{Aumiler2005}%
  \BibitemOpen
  \bibfield  {author} {\bibinfo {author} {\bibfnamefont {D.}~\bibnamefont
  {Aumiler}}, \bibinfo {author} {\bibfnamefont {T.}~\bibnamefont {Ban}},
  \bibinfo {author} {\bibfnamefont {H.}~\bibnamefont
  {Skenderovi\ifmmode~\acute{c}\else \'{c}\fi{}}}, \ and\ \bibinfo {author}
  {\bibfnamefont {G.}~\bibnamefont {Pichler}},\ }\href {\doibase
  10.1103/PhysRevLett.95.233001} {\bibfield  {journal} {\bibinfo  {journal}
  {Phys. Rev. Lett.}\ }\textbf {\bibinfo {volume} {95}},\ \bibinfo {pages}
  {233001} (\bibinfo {year} {2005})}\BibitemShut {NoStop}%
\bibitem [{\citenamefont {Ban}\ \emph {et~al.}(2006)\citenamefont {Ban},
  \citenamefont {Aumiler}, \citenamefont {Skenderovi\ifmmode~\acute{c}\else
  \'{c}\fi{}},\ and\ \citenamefont {Pichler}}]{Ban2006}%
  \BibitemOpen
  \bibfield  {author} {\bibinfo {author} {\bibfnamefont {T.}~\bibnamefont
  {Ban}}, \bibinfo {author} {\bibfnamefont {D.}~\bibnamefont {Aumiler}},
  \bibinfo {author} {\bibfnamefont {H.}~\bibnamefont
  {Skenderovi\ifmmode~\acute{c}\else \'{c}\fi{}}}, \ and\ \bibinfo {author}
  {\bibfnamefont {G.}~\bibnamefont {Pichler}},\ }\href {\doibase
  10.1103/PhysRevA.73.043407} {\bibfield  {journal} {\bibinfo  {journal} {Phys.
  Rev. A}\ }\textbf {\bibinfo {volume} {73}},\ \bibinfo {pages} {043407}
  (\bibinfo {year} {2006})}\BibitemShut {NoStop}%
\bibitem [{\citenamefont {Vuji{\v{c}}i{\'{c}}}\ \emph
  {et~al.}(2007)\citenamefont {Vuji{\v{c}}i{\'{c}}}, \citenamefont
  {Vdovi{\'{c}}}, \citenamefont {Aumiler}, \citenamefont {Ban}, \citenamefont
  {Skenderovi{\'{c}}},\ and\ \citenamefont {Pichler}}]{Vujicic2007}%
  \BibitemOpen
  \bibfield  {author} {\bibinfo {author} {\bibfnamefont {N.}~\bibnamefont
  {Vuji{\v{c}}i{\'{c}}}}, \bibinfo {author} {\bibfnamefont {S.}~\bibnamefont
  {Vdovi{\'{c}}}}, \bibinfo {author} {\bibfnamefont {D.}~\bibnamefont
  {Aumiler}}, \bibinfo {author} {\bibfnamefont {T.}~\bibnamefont {Ban}},
  \bibinfo {author} {\bibfnamefont {H.}~\bibnamefont {Skenderovi{\'{c}}}}, \
  and\ \bibinfo {author} {\bibfnamefont {G.}~\bibnamefont {Pichler}},\ }\href
  {\doibase 10.1140/epjd/e2006-00261-5} {\bibfield  {journal} {\bibinfo
  {journal} {The European Physical Journal D}\ }\textbf {\bibinfo {volume}
  {41}},\ \bibinfo {pages} {447} (\bibinfo {year} {2007})}\BibitemShut
  {NoStop}%
\bibitem [{\citenamefont {Cere}\ \emph {et~al.}(2009)\citenamefont {Cere},
  \citenamefont {Parigi}, \citenamefont {Abad}, \citenamefont {Wolfgramm},
  \citenamefont {Predojević},\ and\ \citenamefont {Mitchell}}]{Cere2009}%
  \BibitemOpen
  \bibfield  {author} {\bibinfo {author} {\bibfnamefont {A.}~\bibnamefont
  {Cere}}, \bibinfo {author} {\bibfnamefont {V.}~\bibnamefont {Parigi}},
  \bibinfo {author} {\bibfnamefont {M.}~\bibnamefont {Abad}}, \bibinfo {author}
  {\bibfnamefont {F.}~\bibnamefont {Wolfgramm}}, \bibinfo {author}
  {\bibfnamefont {A.}~\bibnamefont {Predojević}}, \ and\ \bibinfo {author}
  {\bibfnamefont {M.}~\bibnamefont {Mitchell}},\ }\href {\doibase
  10.1364/OL.34.001012} {\bibfield  {journal} {\bibinfo  {journal} {Optics
  letters}\ }\textbf {\bibinfo {volume} {34}},\ \bibinfo {pages} {1012}
  (\bibinfo {year} {2009})}\BibitemShut {NoStop}%
\bibitem [{\citenamefont {Akulshin}\ \emph {et~al.}(2008)\citenamefont
  {Akulshin}, \citenamefont {Singh}, \citenamefont {Sidorov},\ and\
  \citenamefont {Hannaford}}]{Akulshin08}%
  \BibitemOpen
  \bibfield  {author} {\bibinfo {author} {\bibfnamefont {A.}~\bibnamefont
  {Akulshin}}, \bibinfo {author} {\bibfnamefont {M.}~\bibnamefont {Singh}},
  \bibinfo {author} {\bibfnamefont {A.}~\bibnamefont {Sidorov}}, \ and\
  \bibinfo {author} {\bibfnamefont {P.}~\bibnamefont {Hannaford}},\ }\href
  {\doibase 10.1364/OE.16.015463} {\bibfield  {journal} {\bibinfo  {journal}
  {Opt. Express}\ }\textbf {\bibinfo {volume} {16}},\ \bibinfo {pages} {15463}
  (\bibinfo {year} {2008})}\BibitemShut {NoStop}%
\bibitem [{\citenamefont {Li}\ \emph {et~al.}(2017{\natexlab{a}})\citenamefont
  {Li}, \citenamefont {Peng}, \citenamefont {Budker}, \citenamefont
  {Wickenbrock}, \citenamefont {Pang}, \citenamefont {Zhang},\ and\
  \citenamefont {Guo}}]{Li2017a}%
  \BibitemOpen
  \bibfield  {author} {\bibinfo {author} {\bibfnamefont {W.}~\bibnamefont
  {Li}}, \bibinfo {author} {\bibfnamefont {X.}~\bibnamefont {Peng}}, \bibinfo
  {author} {\bibfnamefont {D.}~\bibnamefont {Budker}}, \bibinfo {author}
  {\bibfnamefont {A.}~\bibnamefont {Wickenbrock}}, \bibinfo {author}
  {\bibfnamefont {B.}~\bibnamefont {Pang}}, \bibinfo {author} {\bibfnamefont
  {R.}~\bibnamefont {Zhang}}, \ and\ \bibinfo {author} {\bibfnamefont
  {H.}~\bibnamefont {Guo}},\ }\href {\doibase 10.1364/OL.42.004163} {\bibfield
  {journal} {\bibinfo  {journal} {Opt. Lett.}\ }\textbf {\bibinfo {volume}
  {42}},\ \bibinfo {pages} {4163} (\bibinfo {year}
  {2017}{\natexlab{a}})}\BibitemShut {NoStop}%
\bibitem [{\citenamefont {Li}\ \emph {et~al.}(2017{\natexlab{b}})\citenamefont
  {Li}, \citenamefont {Balabas}, \citenamefont {Peng}, \citenamefont
  {Pustelny}, \citenamefont {Wickenbrock}, \citenamefont {Guo},\ and\
  \citenamefont {Budker}}]{Li2017b}%
  \BibitemOpen
  \bibfield  {author} {\bibinfo {author} {\bibfnamefont {W.}~\bibnamefont
  {Li}}, \bibinfo {author} {\bibfnamefont {M.}~\bibnamefont {Balabas}},
  \bibinfo {author} {\bibfnamefont {X.}~\bibnamefont {Peng}}, \bibinfo {author}
  {\bibfnamefont {S.}~\bibnamefont {Pustelny}}, \bibinfo {author}
  {\bibfnamefont {A.}~\bibnamefont {Wickenbrock}}, \bibinfo {author}
  {\bibfnamefont {H.}~\bibnamefont {Guo}}, \ and\ \bibinfo {author}
  {\bibfnamefont {D.}~\bibnamefont {Budker}},\ }\href {\doibase
  10.1063/1.4976017} {\bibfield  {journal} {\bibinfo  {journal} {Journal of
  Applied Physics}\ }\textbf {\bibinfo {volume} {121}},\ \bibinfo {pages}
  {063104} (\bibinfo {year} {2017}{\natexlab{b}})}\BibitemShut {NoStop}%
\bibitem [{\citenamefont {Main}\ \emph {et~al.}(2020)\citenamefont {Main},
  \citenamefont {Hird}, \citenamefont {Gao}, \citenamefont {Oguz},
  \citenamefont {Saunders}, \citenamefont {Walmsley},\ and\ \citenamefont
  {Ledingham}}]{Main2020}%
  \BibitemOpen
  \bibfield  {author} {\bibinfo {author} {\bibfnamefont {D.}~\bibnamefont
  {Main}}, \bibinfo {author} {\bibfnamefont {T.~M.}\ \bibnamefont {Hird}},
  \bibinfo {author} {\bibfnamefont {S.}~\bibnamefont {Gao}}, \bibinfo {author}
  {\bibfnamefont {E.}~\bibnamefont {Oguz}}, \bibinfo {author} {\bibfnamefont
  {D.~J.}\ \bibnamefont {Saunders}}, \bibinfo {author} {\bibfnamefont {I.~A.}\
  \bibnamefont {Walmsley}}, \ and\ \bibinfo {author} {\bibfnamefont {P.~M.}\
  \bibnamefont {Ledingham}},\ }\href@noop {} {\bibfield  {journal} {\bibinfo
  {journal} {In preparation}\ } (\bibinfo {year} {2020})}\BibitemShut {NoStop}%
\bibitem [{\citenamefont {Young}\ \emph {et~al.}(1994)\citenamefont {Young},
  \citenamefont {Hill}, \citenamefont {Sibener}, \citenamefont {Price},
  \citenamefont {Tanner}, \citenamefont {Wieman},\ and\ \citenamefont
  {Leone}}]{Young1994}%
  \BibitemOpen
  \bibfield  {author} {\bibinfo {author} {\bibfnamefont {L.}~\bibnamefont
  {Young}}, \bibinfo {author} {\bibfnamefont {W.~T.}\ \bibnamefont {Hill}},
  \bibinfo {author} {\bibfnamefont {S.~J.}\ \bibnamefont {Sibener}}, \bibinfo
  {author} {\bibfnamefont {S.~D.}\ \bibnamefont {Price}}, \bibinfo {author}
  {\bibfnamefont {C.~E.}\ \bibnamefont {Tanner}}, \bibinfo {author}
  {\bibfnamefont {C.~E.}\ \bibnamefont {Wieman}}, \ and\ \bibinfo {author}
  {\bibfnamefont {S.~R.}\ \bibnamefont {Leone}},\ }\href {\doibase
  10.1103/PhysRevA.50.2174} {\bibfield  {journal} {\bibinfo  {journal} {Phys.
  Rev. A}\ }\textbf {\bibinfo {volume} {50}},\ \bibinfo {pages} {2174}
  (\bibinfo {year} {1994})}\BibitemShut {NoStop}%
\bibitem [{\citenamefont {Saunders}\ \emph {et~al.}(2016)\citenamefont
  {Saunders}, \citenamefont {Munns}, \citenamefont {Champion}, \citenamefont
  {Qiu}, \citenamefont {Kaczmarek}, \citenamefont {Poem}, \citenamefont
  {Ledingham}, \citenamefont {Walmsley},\ and\ \citenamefont
  {Nunn}}]{Saunders2016}%
  \BibitemOpen
  \bibfield  {author} {\bibinfo {author} {\bibfnamefont {D.~J.}\ \bibnamefont
  {Saunders}}, \bibinfo {author} {\bibfnamefont {J.~H.~D.}\ \bibnamefont
  {Munns}}, \bibinfo {author} {\bibfnamefont {T.~F.~M.}\ \bibnamefont
  {Champion}}, \bibinfo {author} {\bibfnamefont {C.}~\bibnamefont {Qiu}},
  \bibinfo {author} {\bibfnamefont {K.~T.}\ \bibnamefont {Kaczmarek}}, \bibinfo
  {author} {\bibfnamefont {E.}~\bibnamefont {Poem}}, \bibinfo {author}
  {\bibfnamefont {P.~M.}\ \bibnamefont {Ledingham}}, \bibinfo {author}
  {\bibfnamefont {I.~A.}\ \bibnamefont {Walmsley}}, \ and\ \bibinfo {author}
  {\bibfnamefont {J.}~\bibnamefont {Nunn}},\ }\href {\doibase
  10.1103/PhysRevLett.116.090501} {\bibfield  {journal} {\bibinfo  {journal}
  {Phys. Rev. Lett.}\ }\textbf {\bibinfo {volume} {116}},\ \bibinfo {pages}
  {090501} (\bibinfo {year} {2016})}\BibitemShut {NoStop}%
\bibitem [{\citenamefont {Afzelius}\ and\ \citenamefont
  {Simon}(2010)}]{Afzelius2010b}%
  \BibitemOpen
  \bibfield  {author} {\bibinfo {author} {\bibfnamefont {M.}~\bibnamefont
  {Afzelius}}\ and\ \bibinfo {author} {\bibfnamefont {C.}~\bibnamefont
  {Simon}},\ }\href {\doibase 10.1103/PhysRevA.82.022310} {\bibfield  {journal}
  {\bibinfo  {journal} {Phys. Rev. A}\ }\textbf {\bibinfo {volume} {82}},\
  \bibinfo {pages} {022310} (\bibinfo {year} {2010})}\BibitemShut {NoStop}%
\bibitem [{\citenamefont {Afzelius}\ \emph {et~al.}(2010)\citenamefont
  {Afzelius}, \citenamefont {Usmani}, \citenamefont {Amari}, \citenamefont
  {Lauritzen}, \citenamefont {Walther}, \citenamefont {Simon}, \citenamefont
  {Sangouard}, \citenamefont {Min\'a\ifmmode~\check{r}\else \v{r}\fi{}},
  \citenamefont {de~Riedmatten}, \citenamefont {Gisin},\ and\ \citenamefont
  {Kr\"oll}}]{Afzelius2010}%
  \BibitemOpen
  \bibfield  {author} {\bibinfo {author} {\bibfnamefont {M.}~\bibnamefont
  {Afzelius}}, \bibinfo {author} {\bibfnamefont {I.}~\bibnamefont {Usmani}},
  \bibinfo {author} {\bibfnamefont {A.}~\bibnamefont {Amari}}, \bibinfo
  {author} {\bibfnamefont {B.}~\bibnamefont {Lauritzen}}, \bibinfo {author}
  {\bibfnamefont {A.}~\bibnamefont {Walther}}, \bibinfo {author} {\bibfnamefont
  {C.}~\bibnamefont {Simon}}, \bibinfo {author} {\bibfnamefont
  {N.}~\bibnamefont {Sangouard}}, \bibinfo {author} {\bibfnamefont {J.~c.~v.}\
  \bibnamefont {Min\'a\ifmmode~\check{r}\else \v{r}\fi{}}}, \bibinfo {author}
  {\bibfnamefont {H.}~\bibnamefont {de~Riedmatten}}, \bibinfo {author}
  {\bibfnamefont {N.}~\bibnamefont {Gisin}}, \ and\ \bibinfo {author}
  {\bibfnamefont {S.}~\bibnamefont {Kr\"oll}},\ }\href {\doibase
  10.1103/PhysRevLett.104.040503} {\bibfield  {journal} {\bibinfo  {journal}
  {Phys. Rev. Lett.}\ }\textbf {\bibinfo {volume} {104}},\ \bibinfo {pages}
  {040503} (\bibinfo {year} {2010})}\BibitemShut {NoStop}%
\bibitem [{\citenamefont {Kaczmarek}\ \emph {et~al.}(2018)\citenamefont
  {Kaczmarek}, \citenamefont {Ledingham}, \citenamefont {Brecht}, \citenamefont
  {Thomas}, \citenamefont {Thekkadath}, \citenamefont {{Lazo-Arjona}},
  \citenamefont {Munns}, \citenamefont {Poem}, \citenamefont {Feizpour},
  \citenamefont {Saunders}, \citenamefont {Nunn},\ and\ \citenamefont
  {Walmsley}}]{Kaczmarek2018}%
  \BibitemOpen
  \bibfield  {author} {\bibinfo {author} {\bibfnamefont {K.~T.}\ \bibnamefont
  {Kaczmarek}}, \bibinfo {author} {\bibfnamefont {P.~M.}\ \bibnamefont
  {Ledingham}}, \bibinfo {author} {\bibfnamefont {B.}~\bibnamefont {Brecht}},
  \bibinfo {author} {\bibfnamefont {S.~E.}\ \bibnamefont {Thomas}}, \bibinfo
  {author} {\bibfnamefont {G.~S.}\ \bibnamefont {Thekkadath}}, \bibinfo
  {author} {\bibfnamefont {O.}~\bibnamefont {{Lazo-Arjona}}}, \bibinfo {author}
  {\bibfnamefont {J.~H.~D.}\ \bibnamefont {Munns}}, \bibinfo {author}
  {\bibfnamefont {E.}~\bibnamefont {Poem}}, \bibinfo {author} {\bibfnamefont
  {A.}~\bibnamefont {Feizpour}}, \bibinfo {author} {\bibfnamefont {D.~J.}\
  \bibnamefont {Saunders}}, \bibinfo {author} {\bibfnamefont {J.}~\bibnamefont
  {Nunn}}, \ and\ \bibinfo {author} {\bibfnamefont {I.~A.}\ \bibnamefont
  {Walmsley}},\ }\href {\doibase 10.1103/PhysRevA.97.042316} {\bibfield
  {journal} {\bibinfo  {journal} {Phys. Rev. A}\ }\textbf {\bibinfo {volume}
  {97}},\ \bibinfo {pages} {042316} (\bibinfo {year} {2018})}\BibitemShut
  {NoStop}%
\bibitem [{\citenamefont {Finkelstein}\ \emph {et~al.}(2018)\citenamefont
  {Finkelstein}, \citenamefont {Poem}, \citenamefont {Michel}, \citenamefont
  {Lahad},\ and\ \citenamefont {Firstenberg}}]{Finkelstein2018}%
  \BibitemOpen
  \bibfield  {author} {\bibinfo {author} {\bibfnamefont {R.}~\bibnamefont
  {Finkelstein}}, \bibinfo {author} {\bibfnamefont {E.}~\bibnamefont {Poem}},
  \bibinfo {author} {\bibfnamefont {O.}~\bibnamefont {Michel}}, \bibinfo
  {author} {\bibfnamefont {O.}~\bibnamefont {Lahad}}, \ and\ \bibinfo {author}
  {\bibfnamefont {O.}~\bibnamefont {Firstenberg}},\ }\href {\doibase
  10.1126/sciadv.aap8598} {\bibfield  {journal} {\bibinfo  {journal} {Science
  Advances}\ }\textbf {\bibinfo {volume} {4}},\ \bibinfo {pages} {eaap8598}
  (\bibinfo {year} {2018})}\BibitemShut {NoStop}%
\bibitem [{\citenamefont {Gao}\ \emph {et~al.}(2019)\citenamefont {Gao},
  \citenamefont {Lazo-Arjona}, \citenamefont {Brecht}, \citenamefont
  {Kaczmarek}, \citenamefont {Thomas}, \citenamefont {Nunn}, \citenamefont
  {Ledingham}, \citenamefont {Saunders},\ and\ \citenamefont
  {Walmsley}}]{Gao2019}%
  \BibitemOpen
  \bibfield  {author} {\bibinfo {author} {\bibfnamefont {S.}~\bibnamefont
  {Gao}}, \bibinfo {author} {\bibfnamefont {O.}~\bibnamefont {Lazo-Arjona}},
  \bibinfo {author} {\bibfnamefont {B.}~\bibnamefont {Brecht}}, \bibinfo
  {author} {\bibfnamefont {K.~T.}\ \bibnamefont {Kaczmarek}}, \bibinfo {author}
  {\bibfnamefont {S.~E.}\ \bibnamefont {Thomas}}, \bibinfo {author}
  {\bibfnamefont {J.}~\bibnamefont {Nunn}}, \bibinfo {author} {\bibfnamefont
  {P.~M.}\ \bibnamefont {Ledingham}}, \bibinfo {author} {\bibfnamefont {D.~J.}\
  \bibnamefont {Saunders}}, \ and\ \bibinfo {author} {\bibfnamefont {I.~A.}\
  \bibnamefont {Walmsley}},\ }\href {\doibase 10.1103/PhysRevLett.123.213604}
  {\bibfield  {journal} {\bibinfo  {journal} {Phys. Rev. Lett.}\ }\textbf
  {\bibinfo {volume} {123}},\ \bibinfo {pages} {213604} (\bibinfo {year}
  {2019})}\BibitemShut {NoStop}%
\bibitem [{\citenamefont {Rubio}\ \emph {et~al.}(2018)\citenamefont {Rubio},
  \citenamefont {Viscor}, \citenamefont {Mompart},\ and\ \citenamefont
  {Ahufinger}}]{Rubio2018}%
  \BibitemOpen
  \bibfield  {author} {\bibinfo {author} {\bibfnamefont {J.~L.}\ \bibnamefont
  {Rubio}}, \bibinfo {author} {\bibfnamefont {D.}~\bibnamefont {Viscor}},
  \bibinfo {author} {\bibfnamefont {J.}~\bibnamefont {Mompart}}, \ and\
  \bibinfo {author} {\bibfnamefont {V.}~\bibnamefont {Ahufinger}},\ }\href
  {\doibase 10.1103/PhysRevA.98.043834} {\bibfield  {journal} {\bibinfo
  {journal} {Phys. Rev. A}\ }\textbf {\bibinfo {volume} {98}},\ \bibinfo
  {pages} {043834} (\bibinfo {year} {2018})}\BibitemShut {NoStop}%
\bibitem [{\citenamefont {Adams}\ \emph {et~al.}(2019)\citenamefont {Adams},
  \citenamefont {Pritchard},\ and\ \citenamefont {Shaffer}}]{Adams2019}%
  \BibitemOpen
  \bibfield  {author} {\bibinfo {author} {\bibfnamefont {C.~S.}\ \bibnamefont
  {Adams}}, \bibinfo {author} {\bibfnamefont {J.~D.}\ \bibnamefont
  {Pritchard}}, \ and\ \bibinfo {author} {\bibfnamefont {J.~P.}\ \bibnamefont
  {Shaffer}},\ }\href {\doibase 10.1088/1361-6455/ab52ef} {\bibfield  {journal}
  {\bibinfo  {journal} {Journal of Physics B: Atomic, Molecular and Optical
  Physics}\ }\textbf {\bibinfo {volume} {53}},\ \bibinfo {pages} {012002}
  (\bibinfo {year} {2019})}\BibitemShut {NoStop}%
\bibitem [{\citenamefont {{Fang}}\ \emph {et~al.}(2017)\citenamefont {{Fang}},
  \citenamefont {{Dong}}, \citenamefont {{Meiselman}}, \citenamefont
  {{Cohen}},\ and\ \citenamefont {{Lorenz}}}]{Fang2017}%
  \BibitemOpen
  \bibfield  {author} {\bibinfo {author} {\bibfnamefont {B.}~\bibnamefont
  {{Fang}}}, \bibinfo {author} {\bibfnamefont {S.}~\bibnamefont {{Dong}}},
  \bibinfo {author} {\bibfnamefont {S.}~\bibnamefont {{Meiselman}}}, \bibinfo
  {author} {\bibfnamefont {O.}~\bibnamefont {{Cohen}}}, \ and\ \bibinfo
  {author} {\bibfnamefont {V.~O.}\ \bibnamefont {{Lorenz}}},\ }in\ \href@noop
  {} {\emph {\bibinfo {booktitle} {2017 Conference on Lasers and Electro-Optics
  (CLEO)}}}\ (\bibinfo {year} {2017})\ pp.\ \bibinfo {pages} {1--2}\BibitemShut
  {NoStop}%
\bibitem [{\citenamefont {Nunn}\ \emph {et~al.}(2013)\citenamefont {Nunn},
  \citenamefont {Langford}, \citenamefont {Kolthammer}, \citenamefont
  {Champion}, \citenamefont {Sprague}, \citenamefont {Michelberger},
  \citenamefont {Jin}, \citenamefont {England},\ and\ \citenamefont
  {Walmsley}}]{Nunn2013b}%
  \BibitemOpen
  \bibfield  {author} {\bibinfo {author} {\bibfnamefont {J.}~\bibnamefont
  {Nunn}}, \bibinfo {author} {\bibfnamefont {N.~K.}\ \bibnamefont {Langford}},
  \bibinfo {author} {\bibfnamefont {W.~S.}\ \bibnamefont {Kolthammer}},
  \bibinfo {author} {\bibfnamefont {T.~F.~M.}\ \bibnamefont {Champion}},
  \bibinfo {author} {\bibfnamefont {M.~R.}\ \bibnamefont {Sprague}}, \bibinfo
  {author} {\bibfnamefont {P.~S.}\ \bibnamefont {Michelberger}}, \bibinfo
  {author} {\bibfnamefont {X.-M.}\ \bibnamefont {Jin}}, \bibinfo {author}
  {\bibfnamefont {D.~G.}\ \bibnamefont {England}}, \ and\ \bibinfo {author}
  {\bibfnamefont {I.~A.}\ \bibnamefont {Walmsley}},\ }\href {\doibase
  10.1103/PhysRevLett.110.133601} {\bibfield  {journal} {\bibinfo  {journal}
  {Phys. Rev. Lett.}\ }\textbf {\bibinfo {volume} {110}},\ \bibinfo {pages}
  {133601} (\bibinfo {year} {2013})}\BibitemShut {NoStop}%
\end{thebibliography}

\clearpage


\section*{Supplementary material}

# Room Temperature Atomic Frequency Comb Memory for Light

D. Main[1], T. M. Hird[1,2], S. Gao[1], I. A. Walmsley[1,3], and P. M. Ledingham[1,4]*

[1] *Clarendon Laboratory, University of Oxford, Parks Road, Oxford, OX1 3PU, UK*
[2] *Department of Physics and Astronomy, University College London, London WC1E 6BT, UK*
[3] *QOLS, Department of Physics, Imperial College London, London SW7 2BW, UK*
[4] *Department of Physics and Astronomy, University of Southampton, Southampton SO17 1BJ, UK*

(Dated: August 10, 2020)

## EXPERIMENTAL SETUP

Figure 1 shows a schematic of the experimental setup. Our memory is a paraffin-coated Cs cell of length 25 mm and diameter 10 mm. Three independent narrowband continuous-wave (CW) lasers are utilised. One laser is tuned to the $F = 4 \rightarrow F' = 4$ transition in the $D_2$ line (852 nm - Sacher Lasertechnik ECDL) and is used to perform the initialisation of the ground states. A second laser is tuned to the $F = 3 \rightarrow F' = 4$ transition in the $D_1$ line (895 nm - Toptica DFB) and is used to perform spin-selective optical pumping - see below for more details. The final laser acts to probe the system both in frequency (absorption spectra) and time (AFC echoes) on the $F = 4 \rightarrow F' = 3, 4, 5$ transitions in the $D_2$ line (Toptica DL Pro). Each pump mode has an acousto-optic modulator (AOM) for temporal control of the pumping, while the probe mode has an AOM in a double-pass configuration in the detection. The role of the AOM is two-fold: For the spectroscopy the detection AOM turns on only at the appropriate time (more information beow) for a duration of 100 ns; For the echo experiment the detection AOM serves to protect the single photon detector (more information below) from the AFC preparation. For the creation of the pulses for the echo experiments, a fiber-based electro-optic modulator is used (driven by two arbitrary waveform generators - RF1 and RF2) to create near Gaussian pulses of 2 ns duration. For more details on this pulse carving setup see [1]. We note that this EOM is bypassed when measuring the atomic specta. The pump modes are combined using a dichroic mirror (DM), and then expanded to a beam radius size of 2.7 mm (1.9 mm) for the hyperfine (velocity selective) pump mode before the cell.

We arrange the probe mode (of beam radius 290, mm at the cell) to be counter-propagating the pump modes and a polarising beam splitter (PBS) is used after the cell to separate this mode toward the detection. A bandpass filter (BPF - Thorlabs FL850-10) centered on the D2 laser is placed in the detecion mode to effectively filter the velocity selective pump laser. A silicon avalanche photodetector (Thorlabs PDA120) is used for the spectroscopy and a single-photon avalanche photo-detector (Perkin Elmer SPCM-AQR series) together with a time-tagger module (Swabian Instruments Time Tagger 20) for the echo experiments. For the echo experiments, several neutral density (ND) filters are placed in the pulsed signal mode before the cell to reduce the number of photons per pulse to the few hundreds level, and additional ND is placed in the detection mode to reduce the signal and any leakage to the single photon level.

The atoms physically move out of the probe mode much faster that the laser can be scanned in a mode-hop free manner, therefore it is not possible to probe the entire AFC spectrum after a given velocity-selective pumping procedure. We instead piece together a full spectrum with around a thousand individual probe measurements. The sequence begins with the pump mode open for approximately 998 $\mu$s and power of about 20 mW, emptying the $F = 4$ hyperfine level. The role of the paraffin coating becomes apparent here, assisting in this initialisation by redistributing the velocity classes of the atoms with spin-preserving collisions with the paraffin allowing to use a narrowband laser for efficient pumping. This is followed by a velocity-selective optical pump for a duration of around $\tau_p$ and a varying power level. Then, the probe mode is detected precisely after the velocity-selective pump has turned off for a duration of around 100 ns. The total sequence duration is 1 ms. The probe laser is scanned over about a GHz at a rate of 5 Hz and is scanned asynchronously with the AFC preparation, such that a different part of the AFC spectrum is probed after each preparation round. Repeating this procedure we piece together 1000 probe measurements to complete the full spectrum.

## AFC PREPARATION

As outlined above, the velocity selective pump mode is applied by a single laser tuned to the D1 transition for a duration around 100 ns. To prepare the various atomic frequency combs, it is necessary to address multiple frequency classes. This is done with direct modulation of the laser current via the bias-T with an RF Arbitrary Waveform Generator (Keysight 33600A). Our approach was to use integer divisions of the $F' = 4, F' = 5$ hyperfine excited state splitting of 251 MHz in order to utilise both transitions for a broader band AFC implementation. Modulation on the bias-T of the laser has the effect of inducing frequency sidebands on the spectrum of the laser at plus and minus the modulation frequency, resulting in two additional velocity classes being addressed per applied modulation frequency.

Table I outlines the AFC comb spacings $\Delta$ we prepared, the resulting rephasing times $\tau$, and the re-

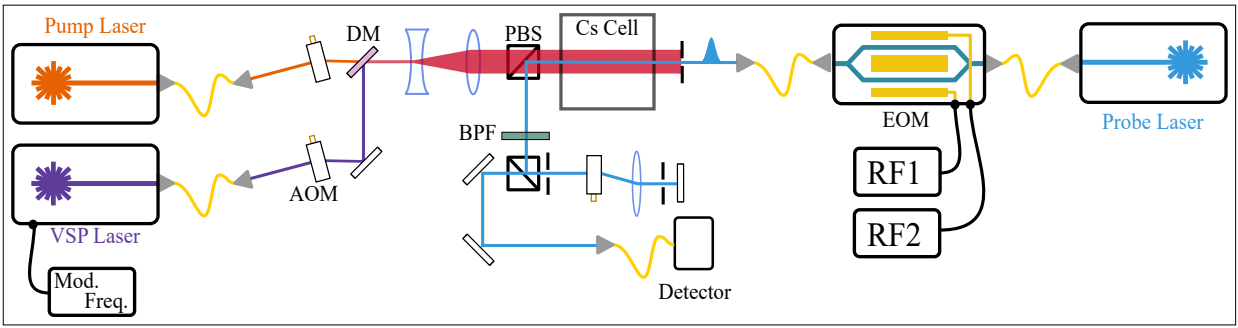

FIG. 1. Experimental setup. See text for details.

quired RF modulation frequencies necessary to create the AFC. Given the maximum analogue output of our waveform generator (80 MHz), it was necessary in some instances to use a half the frequency needed and then to double-amplify-filter the AWG output with mini-circuits electronics. In the instances where two frequencies are required, these where provided by the 2-channel AWG and were combined using a simple coaxial T-piece. The authors note that this approach is far from an ideal implementation involving the use of an RF power splitter, however was sufficient to demonstrate the particular AFC. The RF power of the modulation was optimised for each AFC and was generally not exceeding 6 dBm.

| Δ [MHz] | τ [ns] | Mod Freq. [MHz] |
|---------|--------|-----------------|
| 125.5 | 7.97 | 125.5* |
| 83.67 | 11.95 | 83.67* |
| 62.75 | 15.94 | 62.75, 125.5* |
| 50.2 | 19.92 | 50.2, 100.4* |

TABLE I. AFC tooth separation Δ, corresponding rephasing time τ and the modulation frequencies needed. To create the * frequencies (i.e. frequencies that exceeded the maximum analogue frequency of the AWG), the AWG was set to half the required frequency and a mini-circuits frequency doubler and amplifier were used.

Figure 2 shows the spectra for four separate AFC implementations. Two of these are exactly the combs from the main text with rephasing times of around 8 ns and 12 ns, the additional two showing the case for rephasing times of around 16 ns and 20 ns. This showcases the flexibility of our system and our approach being able to create arbitrary-spaced many-toothed atomic structures within a timescale of a $\mu s$. It is clearly seen that as we decrease the tooth separation Δ, the effective background absorption of the AFC is increasing. This is an expected observation given the

minimum velocity class width that we can prepare, limited by the laser linewidth, power broadening and ultimately the atomic linewidth of the transition [2]. We note that we were not able to observe the time response for the 16 ns and 20 ns rephasing time AFCs, the increased absorption background for these AFCs inducing too much loss to enable efficient read out.

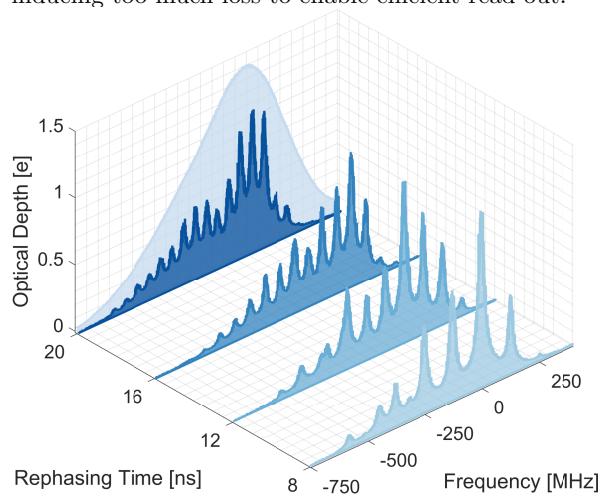

FIG. 2. Different implementations of the AFC. Velocity classes are pumped back into the $F = 4$ hyperfine ground state using the VSP pump laser with modulation frequencies outlined in Table I. The zero of the frequency axis represents the $F = 4$ to $F' = 5$ transition. The unpumped $F = 4$ spectrum is included for reference.

* P.Ledingham@soton.ac.uk

[1] S. E. Thomas, T. M. Hird, J. H. D. Munns, B. Brecht, D. J. Saunders, J. Nunn, I. A. Walmsley, and P. M. Ledingham, Phys. Rev. A **100**, 033801 (2019).

[2] D. Main, T. M. Hird, S. Gao, E. Oguz, D. J. Saunders, I. A. Walmsley, and P. M. Ledingham, In preparation (2020).

 

\end{document}